\documentclass[lettersize,journal]{IEEEtran}
\usepackage{amsmath,amsfonts}
\usepackage{algorithmic}
\usepackage{algorithm}
\usepackage{array}
\usepackage[caption=false,font=normalsize,labelfont=sf,textfont=sf]{subfig}
\usepackage{textcomp}
\usepackage{stfloats}
\usepackage{url}
\usepackage{verbatim}
\usepackage{graphicx}
\usepackage{cite}
\usepackage{amsthm}
\usepackage{tikz}
\usepackage{caption}
\usepackage{float}
\usetikzlibrary {shadows,trees,backgrounds,fit}
\usepackage{tikz-qtree}
\newtheorem{theorem}{Theorem}

\newcommand {\Define} {\stackrel {\Delta} {=}  }
\newcommand{\mya}{\mathrel{\overset{\makebox[0pt]{{\tiny(a)}}}{=}}}
\newcommand{\myb}{\mathrel{\overset{\makebox[0pt]{{\tiny(b)}}}{=}}}
\newcommand{\myc}{\mathrel{\overset{\makebox[0pt]{{\tiny(c)}}}{=}}}

\begin{document}
\title{Zak-OTFS: Pulse Shaping and the Tradeoff between Time/Bandwidth Expansion and Predictability}\author{\IEEEauthorblockN{Jinu Jayachandran\IEEEauthorrefmark{1}, Rahul Kumar Jaiswal\IEEEauthorrefmark{1}, Saif Khan Mohammed\IEEEauthorrefmark{1},
Ronny Hadani\IEEEauthorrefmark{2},
Ananthanarayanan Chockalingam\IEEEauthorrefmark{3}, and 
Robert Calderbank\IEEEauthorrefmark{4},~\IEEEmembership{Fellow,~IEEE}}\\
\IEEEauthorblockA{\IEEEauthorrefmark{1}Department of Electrical Engineering, Indian Institute of Technology Delhi, India}\\
\IEEEauthorblockA{\IEEEauthorrefmark{2}Department of Mathematics, University of Texas at Austin, USA}\\
\IEEEauthorblockA{\IEEEauthorrefmark{3} Department of Electrical Communication Engineering, Indian Institute of Science Bangalore, India}\\
\IEEEauthorblockA{\IEEEauthorrefmark{4}Department of Electrical and Computer Engineering, Duke University, USA}
\thanks{This work has been submitted to the IEEE for possible publication. Copyright may be transferred without notice, after which this version may no longer be accessible.}
\thanks{S. K. Mohammed is also associated with the Bharti School of Telecom. Tech. and Management, IIT Delhi. The work of Saif Khan Mohammed was supported in part by a project (at BSTTM) sponsored by Bharti Airtel Limited India, and in part by the Jai Gupta Chair at IIT Delhi.}
}




    \tikzstyle{flowchart}=[rectangle,minimum height=1cm, minimum width=5cm, text centered, text width=4cm]
    \tikzstyle{roundedrect}=[rounded corners, rectangle,minimum height=1.5cm, minimum width=1cm, text centered, text width=4cm]
    \tikzstyle{treerect}=[rounded corners, rectangle,minimum height=1cm, minimum width=3cm, text centered, text width=3cm]
    \tikzstyle{arrow} = [thick,->,>=stealth]
    \tikzstyle{bigbox} = [draw, dashed,  rounded corners, rectangle,minimum height=2cm,minimum width=5cm]

\maketitle

\begin{abstract}
The Zak-OTFS input/output (I/O) relation is predictable and non-fading when the delay and Doppler periods are greater than the effective channel delay and Doppler spreads, a condition which we refer to as the crystallization condition. When the crystallization condition is satisfied, we describe how to integrate sensing and communication within a single Zak-OTFS subframe by transmitting a pilot in the center of the subframe and surrounding the pilot with a pilot region and guard band to mitigate interference between data symbols and pilot. At the receiver we first read off the effective channel taps within the pilot region, and then use the estimated channel taps to recover the data from the symbols received outside the pilot region. We introduce a framework for filter design in the delay-Doppler (DD) domain where the symplectic Fourier transform connects aliasing in the DD domain (predictability of the I/O relation) with time/bandwidth expansion. The choice of pulse shaping filter determines the fraction of pilot energy that lies outside the pilot region and the degradation in BER performance that results from the interference to data symbols. We demonstrate that Gaussian filters in the DD domain provide significant improvements in BER performance over the sinc and root raised cosine filters considered in previous work. We also demonstrate that, by limiting DD domain aliasing, Gaussian filters extend the region where the crystallization condition is satisfied. The Gaussian filters considered in this paper are a particular case of factorizable pulse shaping filters in the DD domain, and this family of filters may be of independent interest.
\end{abstract}


\begin{figure*}
\vspace{-7mm}
\centering
\includegraphics[width=15cm,height=8.5cm]{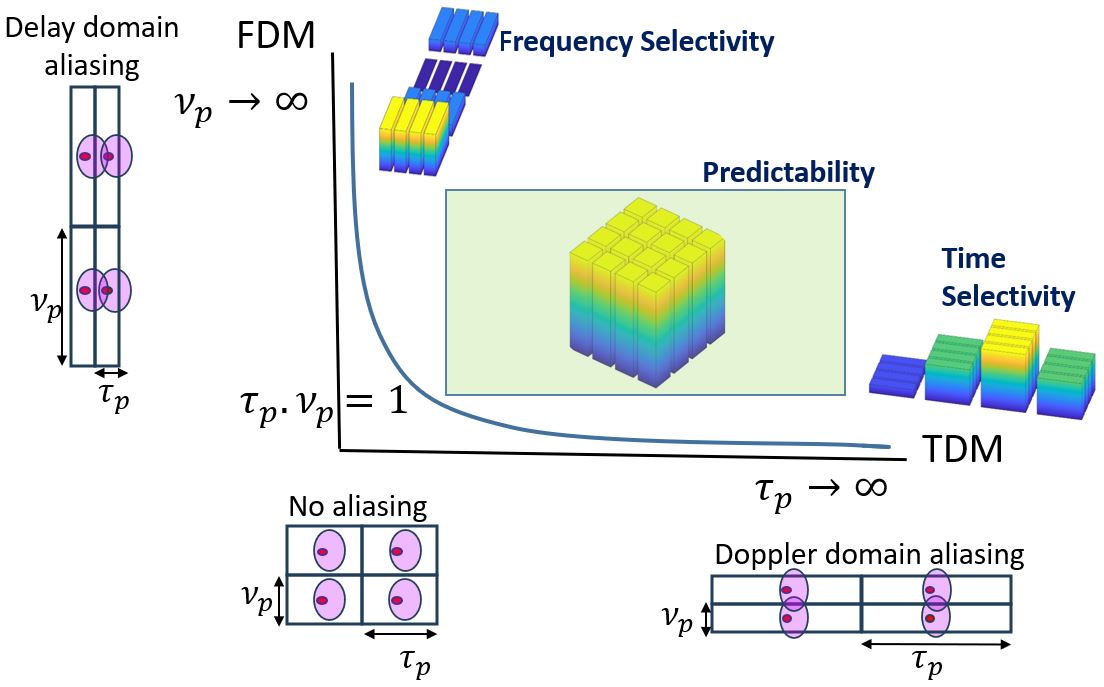}
\caption{Non-predictability results from aliasing in the DD domain illustrated as overlap between the quasi-periodic aliases of the received pulse (whose support are depicted as ellipses). Aliasing occurs when the effective channel spreads are larger than the delay-Doppler periods.
} 
\label{fig_aliasing}
\end{figure*}

\vspace{-4mm}
\section{Introduction}
{ 6G propagation environments are expected to be challenging in terms of extreme Dopplers (few KHz to tens of KHz) for which communication waveforms that are robust in  extreme time selective channels are crucial \cite{intro1},\cite{intro2}.}
A time-domain (TD) pulse is an ideal waveform for pure delay channels as it is possible to separate reflections according to their range. Similarly, a frequency domain (FD) pulse is an ideal waveform for pure Doppler channels since it is possible to separate reflections according to their velocity. The Zak-OTFS carrier is a pulse in the delay-Doppler (DD) domain, and it is designed to separate reflections in both range and velocity, making it a natural fit to doubly spread channels {\cite{derotfs}-\cite{zakotfs2}}. A pulse in the DD domain is a quasi-periodic localized function, defined by a delay period $\tau_p$ and a Doppler period $\nu_p$, where $\tau_p \, \nu_p = 1$. When viewed in the time domain, it is realized as a pulse train modulated by a tone (hence the name \emph{pulsone}). Section \ref{sysmodelsec} describes the Zak-OTFS system model.

We have shown in \cite{zakotfs1}, \cite{zakotfs2} that the Zak-OTFS input/output (I/O) relation is predictable and non-fading when the delay period $\tau_p$ is greater than the effective channel delay spread, and the Doppler period $\nu_p$ is greater than the effective channel Doppler spread. We refer to this condition as the \emph{crystallization condition} with respect to the \emph{period lattice} $\Lambda_p$ comprising integer linear combinations of $(\tau_p,0)$ and $(0,\nu_p)$. Fig.~\ref{fig_aliasing} illustrates how non-predictability and fading result from aliasing in the DD domain, and why the crystallization condition prevents aliasing (see Section II in \cite{zakotfs2}). 

As the Doppler period shrinks, aliasing in Doppler results in time selectivity. As the delay period shrinks, aliasing in delay results in frequency selectivity. When the crystallization condition is satisfied there is no aliasing, the received power profile in the DD domain is flat (there is no fading), and the I/O relation is predictable. 
%


Why Zak-OTFS rather than OFDM? Perhaps the most important reason is that 6G propagation environments are changing the balance between time-frequency methods focused on OFDM signal processing and delay-Doppler methods focused on Zak-OTFS signal processing. OFDM signals live on a coarse grid, and cyclic prefix/carrier spacing are designed to prevent inter carrier interference (ICI). When there is no ICI, equalization in OFDM is relatively simple once the I/O relation is acquired. However, acquisition of the I/O relation in OFDM is non-trivial and model-dependent {\cite{ofdm1}}. By contrast, Zak-OTFS signals live on a fine grid and Zak-OTFS signal processing embraces inter-carrier interference, hence equalization is more involved. {However, when the crystallization condition holds, the I/O relation is predictable, in that the effective channel taps can be read off from the response to a single pilot signal (see \cite{zakotfs1}, \cite{zakotfs2})}. 
{The prediction accuracy of the I/O relation is determined by how slow/fast the DD domain transmit/receive pulse shaping filter response decays \cite{zakotfs2}. A slower decay implies a larger aliasing residue, resulting in a larger prediction error.} 
By focusing not on acquiring the channel, but on acquiring the interaction of channel and modulation, Zak-OTFS circumvents the standard channel-model-dependent approach to wireless communication and operates \emph{model-free}.

Acquiring the I/O relation becomes more critical in 6G propagation environments as Doppler spreads measured in KHz make it more and more difficult to prevent ICI and to estimate channels. Most challenging is the combination of unresolvable paths and high channel spreads. In this paper, we focus on the Veh-A channel {\cite{EVAITU}} specified in Table-\ref{tab_veha}, first because it is representative of real propagation environment, and second because it is very difficult to make the model-dependent mode of operation work. For example, given a channel bandwidth $B=0.96$ MHz, the delay shifts introduced by the first three paths ($0, 0.31$ and $0.71$ $\mu s$) are less than the delay domain resolution $1/B = 1.04 \, \mu s$  and are not resolvable. In these environments, Zak-OTFS enables a model-free approach with superior performance (see \cite{zakotfs2} for details).
      \begin{table}[]
            \centering
            \caption{Power-delay profile of Veh-A channel model}
            \begin{tabular}{|c|c|c|c|c|c|c|}
                \hline
                Path number ($i$)       &  1 &  2  &  3  &  4  &  5  &  6  \\
                \hline
                $\tau_i$ ($\mu$s) &  0 & 0.31& 0.71& 1.09 & 1.73 & 2.51 \\
                \hline
                Relative power ($p_i$) dB   &  0 & -1  & -9  & -10 & -15 & -20\\
                \hline
            \end{tabular}
            \label{tab_veha}
            \vspace{2mm}
            
            \vspace{-4mm}
        \end{table}
%
%
We now highlight our main contributions.

\textbf{Integrated sensing and communication:}
Prior work \cite{zakotfs2} dedicates separate Zak-OTFS subframes to sensing and data communication. We integrate sensing and communication by introducing
an embedded pilot frame\footnote{{Point pilot and embedded pilot schemes have been considered for DD channel estimation in multicarrier OTFS (MC-OTFS) that performs time domain to DD domain conversion (and vice versa) in two steps \cite{mcotfs1}-\cite{embed2}. Zak-OTFS performs this conversion in one step using Zak transform. It has been shown in \cite{zakotfs2} that Zak-OTFS performs better than MC-OTFS over a larger range of Doppler spreads, which is attractive for 6G.}}, where we transmit a pilot in the center of the subframe and surround the pilot with a pilot region for acquiring the taps of the effective channel filter and a guard band to mitigate the interference between the data symbols and pilot.
Thus, we integrate sensing and communication by dividing DD domain resources between the two objectives, and the pilot region and guard band are an overhead that reduce the effective throughput. At the receiver we first read off the effective channel taps from the pilot response received within the pilot region. We then use the estimated channel taps to recover the data from the symbols received outside the pilot region. 

\textbf{Framework for filter design:}
For LTI channels, the Fourier transform connects
localization of pulse shape in time with bandwidth expansion.
For LTV channels, we introduce a framework for filter design where the symplectic Fourier transform
connects localization of pulse shape in the DD domain with time/bandwidth expansion. 
Prior work on MC-OTFS (\cite{mcpulse1}, \cite{mcpulse2}, \cite{ddpulse1}) employs time-domain filters, whereas we employ transmit/receive filters that are DD domain filters. An advantage of our approach is that we are able to represent the effective DD channel as a simple cascade of twisted convolutions (\cite{zakotfs1}, \cite{zakotfs2}).

\begin{figure}
\centering
\includegraphics[width=9cm,height=8cm]{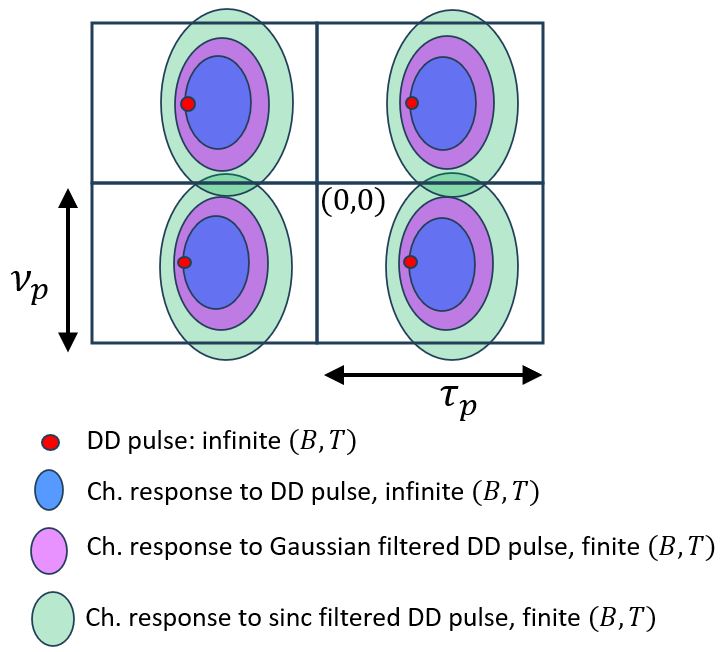}
\caption{The impact of pulse shaping on the support of the effective channel. When the effective channel spread exceeds a DD domain period, aliasing leads to non-predictability.
Here, the effective channel spread for the sinc pulse is larger than that for the Gaussian pulse (green vs. purple ellipses), resulting in Doppler domain aliasing and therefore non-predictability with the sinc pulse. There is no aliasing with the Gaussian pulse, hence the channel interaction is predictable.
} 
\label{efffilter}
\end{figure}

{ Fig.~\ref{efffilter} illustrates how the choice of transmit/receive pulse shaping filter determines the spread of the received pilot response in the DD domain. The DD domain spread of the received pilot response is equal to the spread
of the effective channel filter. A higher spread degrades BER performance since, (i) it results in higher degree of aliasing between the
quasi-periodic aliases of the received pilot response (shown as ellipses with the same color in Fig.~\ref{efffilter}), which degrades the accuracy of the acquired taps of the effective channel filter
and, (ii) a higher spread also results in higher interference between the data symbols and the pilot symbol.
Since the effective channel filter is the twisted convolution of the physical channel filter with the pulse shaping filters, its spread is simply the sum of the spread of the physical channel filter and the spread of the pulse shaping filters.
The time and bandwidth of a Zak-OTFS subframe are inversely related to the Doppler and delay spread respectively of the transmit pulse shaping filter. Therefore, time and bandwidth expansion reduces the Doppler and delay spread respectively, which reduces the spread of the effective channel filter. This in turn improves error rate performance, but at the cost of lower throughput. Therefore, ``good" transmit pulse shaping filters are those which have the smallest possible spread for a given time/bandwidth. Due to the exponential roll-off of Gaussian pulses, Gaussian pulse shaping filters achieve a good trade-off between DD domain spreading and bandwidth expansion.}   

 Section \ref{sec3} introduces Gaussian filters in the DD domain\footnote{\footnotesize{However, we note that P. M. Woodward, in his 1953 book \cite{PMWoodward}, proposed a radar waveform comprising a train of narrow TD Gaussian pulses modulated by a broad Gaussian envelope. This structure appears in the TD when a Gaussian pulse shaping filter is applied to the DD domain pulse located at the origin of the information grid.}}, and in Section \ref{simsecpaper4} we demonstrate significant improvements in BER performance over the sinc and root raised cosine (RRC) filters in the embedded pilot framework.

The reason for the improvement is that outside the main lobe, the magnitude of the Gaussian pulse is significantly lower than the sinc and RRC pulses. In Section \ref{simsecpaper4} we show that the choice of filter determines how BER performance varies with the ratio of pilot power to data power (PDR). For the sinc and RRC filters, the BER curve exhibits a ``U'' shape. The pilot used for sensing interferes with Zak-OTFS carriers used for data transmission. We observe a PDR threshold below which accuracy of sensing dominates BER performance, so that as PDR increases the BER performance improves. Above this PDR threshold interference from the pilot dominates BER performance, so that as PDR increases the BER performance degrades. For Gaussian filters there is very little leakage outside the main lobe,  the interference from the pilot is very small, and for PDRs of interest it does not have a significant impact on BER performance. The crystallization condition applies to the effective channel which depends on the choice of filter. Zak-OTFS is a family of modulations parameterized by the period curve $\tau_p \, \nu_p = 1$ and we become subject to aliasing in the delay-Doppler domain as we shrink the delay period $\tau_p$ or the Doppler period $\nu_p$. Gaussian filters limit aliasing in the delay-Doppler domain, and this translates to increasing the length of the segment of the period curve where the crystallization conditions are satisfied. In Section \ref{simsecpaper4} we show that the increase in length with respect to the sinc pulse is significant.\\

The design of factorizable pulse shaping filters in the delay-Doppler domain may be of independent interest. First, the time and bandwidth properties of the carrier waveforms follow naturally from the Fourier properties of the factors. Second, the use of a matched filter at the receiver makes it possible to compute the effective discrete DD domain channel filter in closed form.

\section{Zak-OTFS System Model}
\label{sysmodelsec}
\begin{figure*}
\vspace{-7mm}
\centering
\includegraphics[width=16.5cm, height=4.5cm]{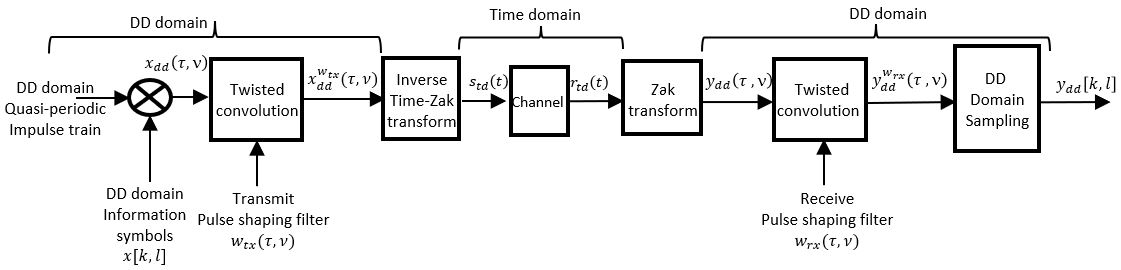}
\caption{Zak-OTFS transceiver processing.} 
\label{figzakotfspaper2}
\end{figure*}

We consider a channel with $P$ discrete paths,
where the $i$-th path has gain $h_i$, delay shift $\tau_i$, and Doppler shift $\nu_i$. The delay-Doppler spreading function $h_{\mbox{\scriptsize{phy}}}(\tau, \nu)$ is given by
\begin{eqnarray}
 \label{eqnp2}
h_{\mbox{\scriptsize{phy}}}(\tau, \nu) & = & \sum\limits_{i=1}^P h_i \, \delta(\tau - \tau_i) \, \delta(\nu - \nu_i).
 \end{eqnarray}
The received TD signal is then given by

{\vspace{-4mm}
\small
 \begin{eqnarray}
r_{\mbox{\scriptsize{td}}}(t) & \hspace{-2.5mm}  = &   \hspace{-2.5mm} \iint h_{\mbox{\scriptsize{phy}}}(\tau, \nu) \, s_{\mbox{\scriptsize{td}}}(t - \tau) \, e^{j 2 \pi \nu (t - \tau)} \, d\tau \, d\nu \, + \, n(t),
 \end{eqnarray}\normalsize}where $s_{\mbox{\scriptsize{td}}}(t)$ is the transmitted TD signal and $n(t)$ is additive white Gaussian noise (AWGN) with PSD $N_0$ W/Hz (see \cite{Bello63}).

\subsection{Zak-OTFS transceiver processing}
\label{sec2apaper4}
In this Section we describe Zak-OTFS transceiver processing as illustrated in Fig.~\ref{figzakotfspaper2} (see Section II of \cite{zakotfs2} for more details). A pulse in the DD domain is a quasi-periodic localized function defined by a delay period $\tau_p$ and a Doppler period $\nu_p = 1/\tau_p$. The TD Zak-OTFS subframe is limited to a time duration $T = N \tau_p$ and a bandwidth $B = M \nu_p$. We transmit $MN$ symbols $x[k,l]$ in each subframe, indexed by $k=0,1,\cdots, M-1$, $l=0,1,\cdots, N-1$. The symbol $x[k,l]$ is carried by the DD domain pulse $x_{\mbox{\scriptsize{dd}}}[k,l]$ given by
 \begin{eqnarray}
 \label{eqn3p4}
     x_{\mbox{\scriptsize{dd}}}[k,l] & = & e^{j 2 \pi \frac{\left\lfloor \frac{k}{M} \right\rfloor l}{N}} \, x[k \, \mbox{\small{mod}} \, M \, , \, l \, \mbox{\small{mod}} \, N]
 \end{eqnarray}for all $k,l \in {\mathbb Z}$. The discrete DD domain pulse $x_{\mbox{\scriptsize{dd}}}[k,l]$ is a \emph{quasi-periodic} function with period $M$ along the delay axis and period $N$ along the Doppler axis. For any $n,m \in {\mathbb Z}$ we have
 \begin{eqnarray}
     \label{qpeqn}
     x_{\mbox{\scriptsize{dd}}}[k + nM,l + mN] & = & e^{j 2 \pi \frac{n l}{N}} \, x_{\mbox{\scriptsize{dd}}}[k,l].
 \end{eqnarray}We note that the discrete DD domain signals $x_{\mbox{\scriptsize{dd}}}[k,l]$ are supported on the \emph{information lattice} $\Lambda_{dd} = \{\left( \frac{k \tau_p}{M} , \frac{l \nu_p}{N} \right) \, | \, k, l \in {\mathbb Z} \}$.

 Next we lift the discrete signal $x_{\mbox{\scriptsize{dd}}}[k,l]$ to the continuous signal
 \begin{eqnarray}
 \label{eqn2}
     x_{\mbox{\scriptsize{dd}}}(\tau, \nu) & = & \sum\limits_{k,l \in {\mathbb Z}} x_{\mbox{\scriptsize{dd}}}[k,l] \, \delta\left(\tau -  \frac{k \tau_p}{M} \right) \, \delta\left(\nu -  \frac{l \nu_p}{N} \right).
 \end{eqnarray}Note that, for any $n,m \in {\mathbb Z}$
 \begin{eqnarray}
     x_{\mbox{\scriptsize{dd}}}(\tau + n \tau_p, \nu + m \nu_p) & = & e^{j 2 \pi n \nu \tau_p} \, x_{\mbox{\scriptsize{dd}}}(\tau, \nu),
 \end{eqnarray}so that $x_{\mbox{\scriptsize{dd}}}(\tau, \nu)$ is periodic with period $\nu_p$ along the Doppler axis and quasi-periodic with period $\tau_p$ along the delay axis. 
 
  The DD domain transmit signal $x_{\mbox{\scriptsize{dd}}}^{w_{tx}}(\tau, \nu)$ is given by the twisted convolution of the transmit pulse shaping filter $w_{tx}(\tau, \nu)$
  with $x_{\mbox{\scriptsize{dd}}}(\tau, \nu)$. 
 \begin{eqnarray}
     x_{\mbox{\scriptsize{dd}}}^{w_{tx}}(\tau, \nu) & = & w_{tx}(\tau, \nu) \, *_{\sigma} \, x_{\mbox{\scriptsize{dd}}}(\tau, \nu) \nonumber \\
     &  &  \hspace{-24mm} = \iint \hspace{-1.5mm} w_{tx}(\tau', \nu') \, x_{\mbox{\scriptsize{dd}}}(\tau - \tau', \nu - \nu') \, e^{j2 \pi \nu' (\tau - \tau')} \, d\tau' \, d\nu',
 \end{eqnarray}where $*_{\sigma}$ denotes twisted convolution. The transmitted TD signal $s_{\mbox{\scriptsize{td}}}(t)$ is the TD realization of $x_{\mbox{\scriptsize{dd}}}^{w_{tx}}(\tau, \nu)$
 given by
 \begin{eqnarray}
 \label{eqn8p4}
 s_{\mbox{\scriptsize{td}}}(t) & = & {\mathcal Z}_t^{-1}\left( x_{\mbox{\scriptsize{dd}}}^{w_{tx}}(\tau, \nu) \right),
 \end{eqnarray}where ${\mathcal Z}_t^{-1}$ denotes the inverse Zak transform (see Eqn.~$(7)$ in \cite{zakotfs1} for more details). Pulse shaping is required to limit the time and bandwidth of $s_{\mbox{\scriptsize{td}}}(t)$. In the absence of pulse shaping (i.e., $w_{tx}(\tau, \nu) = \delta(\tau, \nu)$) the transmit signal has infinite duration and bandwidth.

  At the receiver, we pass from the TD to the DD domain by applying the Zak transform ${\mathcal Z}_t$ to the received time domain signal $r_{\mbox{\scriptsize{td}}}(t)$, and we obtain
 \begin{eqnarray}
     y_{\mbox{\scriptsize{dd}}}(\tau, \nu) & = & {\mathcal Z}_t\left( r_{\mbox{\scriptsize{td}}}(t) \right).
 \end{eqnarray}Next we apply a matched filter $w_{rx}(\tau, \nu)$ which acts by twisted convolution on
 $y_{\mbox{\scriptsize{dd}}}(\tau, \nu)$ to give
 \begin{eqnarray}
y_{\mbox{\scriptsize{dd}}}^{w_{rx}}(\tau, \nu) & = & w_{rx}(\tau, \nu) \, *_{\sigma}\, y_{\mbox{\scriptsize{dd}}}(\tau, \nu).
 \end{eqnarray}The filtered signal $y_{\mbox{\scriptsize{dd}}}^{w_{rx}}(\tau, \nu)$ is then sampled on the information lattice $\Lambda_{dd}$, resulting in the
 discrete quasi-periodic DD domain received signal
 $y_{\mbox{\scriptsize{dd}}}[k,l]$ given by
 \begin{eqnarray}
 \label{eqn9}
     y_{\mbox{\scriptsize{dd}}}[k,l] & = & y_{\mbox{\scriptsize{dd}}}^{w_{rx}}\left( \frac{k \tau_p}{M} \,,\, \frac{l \nu_p}{N}  \right),
 \end{eqnarray}for $k,l \in {\mathbb Z}$.
 Equations (\ref{eqn2}) to (\ref{eqn9}) combine to give the I/O relation \cite{zakotfs2}
 \begin{eqnarray}
 \label{zakotfsiorel}
y_{\mbox{\scriptsize{dd}}}[k,l] & = & h_{\mbox{\scriptsize{eff}}}[k,l] \, *_{\sigma} \, x_{\mbox{\scriptsize{dd}}}[k,l] \, + \, n_{\mbox{\scriptsize{dd}}}[k,l],
 \end{eqnarray}where $*_{\sigma}$ becomes twisted convolution in the discrete DD domain given by
 \begin{eqnarray}
h_{\mbox{\scriptsize{eff}}}[k,l] \, *_{\sigma} \, x_{\mbox{\scriptsize{dd}}}[k,l] & & \nonumber \\
& &  \hspace{-30mm} =  \sum\limits_{k', l' \in {\mathbb Z}} h_{\mbox{\scriptsize{eff}}}[k',l']  \,  x_{\mbox{\scriptsize{dd}}}[k -k',l - l'] \, e^{j 2 \pi \frac{l' (k - k')}{MN}}.
 \end{eqnarray}The effective channel filter $h_{\mbox{\scriptsize{eff}}}[k,l]$ is given by
 \begin{eqnarray}
 \label{eqn13p5}
     h_{\mbox{\scriptsize{eff}}}[k,l] & = & h_{\mbox{\scriptsize{eff}}}\left( \tau = \frac{k \tau_p}{M}, \nu = \frac{l \nu_p}{N} \right),
     \end{eqnarray}where
     \begin{eqnarray}
     \label{eqnhefftwisted}
     h_{\mbox{\scriptsize{eff}}}\left( \tau,\nu \right) & = & w_{rx}(\tau, \nu) *_{\sigma} h_{\mbox{\scriptsize{phy}}}(\tau, \nu) *_{\sigma}  w_{tx}(\tau, \nu).
 \end{eqnarray}
 The I/O relation in (\ref{zakotfsiorel})
 is said to be predictable if the channel response $h_{\mbox{\scriptsize{eff}}}[k,l] \, *_{\sigma} \, x_{\mbox{\scriptsize{dd,1}}}[k,l]$ to any arbitrary input $x_{\mbox{\scriptsize{dd,1}}}[k,l]$ can be predicted from the channel response to a known input \cite{zakotfs1, zakotfs2}. The I/O relation is predictable if the crystallization condition is satisfied, i.e., if the delay spread $\tau_{ds}$ and Doppler spread $\nu_{ds}$ of $h_{\mbox{\scriptsize{eff}}}\left( \tau,\nu \right)$ are less than the delay and Doppler period respectively \cite{zakotfs1}, \cite{zakotfs2}, i.e.,
 \begin{eqnarray}
 \label{eqncryscnd}
     \tau_p & > & \tau_{ds} \,\,\mbox{\small{and}}\,\, \nu_p \, > \, \nu_{ds}.
 \end{eqnarray}In the absence of any time/bandwidth constraint (i.e., infinite time/bandwidth), no pulse shaping is required (i.e., $w_{rx}(\tau, \nu) = w_{rx}(\tau, \nu) = \delta(\tau, \nu)$) and $h_{\mbox{\scriptsize{eff}}}\left( \tau,\nu \right) = h_{\mbox{\scriptsize{phy}}}\left( \tau,\nu \right)$. In the absence of pulse shaping,
 the I/O relation is predictable when the delay and Doppler spread of the physical channel spreading function/filter $h_{\mbox{\scriptsize{phy}}}\left( \tau,\nu \right)$ is less than the delay and Doppler period respectively.

 The samples $y_{\mbox{\scriptsize{dd}}}[k,l]$ are quasi-periodic, and it follows from (\ref{qpeqn}) that the $MN$ information symbols are determined by the $MN$ samples $y_{\mbox{\scriptsize{dd}}}[k,l]$, $k=0,1,\cdots, M-1$, $l=0,1,\cdots, N-1$. We write the samples ${y}_{\mbox{\scriptsize{dd}}}[k,l]$ as a vector ${\bf y}_{\mbox{\scriptsize{dd}}} \in {\mathbb C}^{MN}$, and we write the I/O relation in matrix-vector form as
 \begin{eqnarray}
 \label{eqn87624}
     {\bf y}_{\mbox{\scriptsize{dd}}} & = & {\bf H}_{\mbox{\scriptsize{dd}}} \, {\bf x}_{\mbox{\scriptsize{dd}}}  \, + \, {\bf n}_{\mbox{\scriptsize{dd}}},
 \end{eqnarray}where we have written the $MN$ information symbols $x[k,l]$ as a vector ${\bf x}_{\mbox{\scriptsize{dd}}} \in {\mathbb C}^{MN}$, and the $MN$ noise samples $n_{\mbox{\scriptsize{dd}}}[k,l]$, $k=0,1,\cdots, M-1$, $l=0,1,\cdots, N-1$ as a vector ${\bf n}_{\mbox{\scriptsize{dd}}} \in {\mathbb C}^{MN}$ (see $(39)$ in\cite{zakotfs2}). The entries ${ H}_{\mbox{\scriptsize{dd}}}[k'N + l', kN + l]$, $k',k = 0,1,\cdots, M-1$, $l',l =0,1,\cdots, N-1$ of the effective channel matrix ${\bf H}_{\mbox{\scriptsize{dd}}} \in {\mathbb C}^{MN \times MN}$ are given by (\ref{paper2_eqn12}) (see top of this page). 

 \begin{figure*}
\vspace{-8mm}
{\small
\begin{eqnarray}
\label{paper2_eqn12}
{ H}_{_{\mbox{\footnotesize{dd}}}}[k'N + l' , k N + l ]  & = & \sum\limits_{n=-\infty}^{\infty} \sum\limits_{m = -\infty}^{\infty} h_{_{\mbox{\footnotesize{eff}}}}[k' - k -nM, l' - l -mN] \, e^{j 2 \pi n l/N} e^{j2 \pi \frac{l' - l - mN}{N} \frac{k + nM}{M}}, \nonumber \\
& & k',k=0,1,\cdots, M-1, \,\, l',l=0,1,\cdots, N-1.
\end{eqnarray}\normalsize}
\vspace{-5mm}
\begin{eqnarray*}
    \hline
\end{eqnarray*}
\vspace{-4mm}
\end{figure*}We now consider a factorizable transmit pulse
shaping filter that takes the form 
\begin{eqnarray}
    w_{tx}(\tau, \nu) = w_1(\tau) \, w_2(\nu).
\end{eqnarray}The transmitted TD signal $s_{\mbox{\scriptsize{td}}}(t)$ is the superposition of carrier waveforms $s_{\mbox{\scriptsize{td}},k,l}(t)$ modulated by information symbols $x[k,l]$ (see Appendix \ref{appCp4} for derivation):
\begin{eqnarray}
\label{paper4carrierwaveforms}
    s_{\mbox{\scriptsize{td}}}(t) & \hspace{-2mm} = & \hspace{-2mm} \sum\limits_{k=0}^{M-1} \sum\limits_{l=0}^{N-1} x[k,l] \, s_{\mbox{\scriptsize{td}},k,l}(t).
\end{eqnarray}The $(k,l)$-th carrier waveform $s_{\mbox{\scriptsize{td}},k,l}(t)$ is given by
\begin{eqnarray}
    \label{eqn19prev}
    s_{\mbox{\scriptsize{td}},k,l}(t) & \hspace{-3mm} = & \hspace{-3mm} \iint w_1(\tau) \, w_2(\nu) \, x_{k,l}(t - \tau) \, e^{j 2 \pi \nu (t - \tau)} \, d\tau \, d\nu \nonumber \\
\end{eqnarray}where
\begin{eqnarray}
x_{k,l}(t) & = & \sqrt{\tau_p} \sum\limits_{n \in {\mathbb Z}} e^{j 2 \pi \frac{n l}{N} } \, \delta\left(t - n \tau_p -  \frac{k \tau_p}{M} \right).
\end{eqnarray}We rewrite (\ref{eqn19prev}) as
\begin{eqnarray}
\label{eqn20prev}
    s_{\mbox{\scriptsize{td}},k,l}(t) & = &  \int w_1(\tau) \, x_{k,l}(t - \tau) \, W_2(t - \tau) \, d\tau \nonumber \\
    & = & w_1(t) \, \star \, \left[ W_2(t) \, x_{k,l}(t) \right],
\end{eqnarray}where $\star$ denotes linear convolution and
\begin{eqnarray}
    W_2(t) & = & \int w_2(\nu) \, e^{j 2 \pi \nu t} \, d\nu
\end{eqnarray}is the inverse Fourier transform of the factor $w_2(\nu)$. Since $x_{k,l}(t)$ has infinite time duration,
the time duration $T'$ of each Zak-OTFS carrier (and that of the Zak-OTFS subframe) is simply the time duration of $W_2(t)$. Hence the Doppler axis spread of $w_2(\nu)$ is approximately $1/T'$.

We now set
\begin{eqnarray}
x_{2,k,l}(t) & = & W_2(t) \, x_{k,l}(t).
\end{eqnarray}It follows from (\ref{eqn20prev}) that the
Fourier transform of $s_{\mbox{\scriptsize{td}},k,l}(t)$
is given by
\begin{eqnarray}
    s_{\mbox{\scriptsize{fd}},k,l}(f) & = & W_1(f) \, X_{2,k,l}(f),
\end{eqnarray}where $W_1(f)$ is the Fourier transform of $w_1(t)$ and $X_{2,k,l}(f)$ is the Fourier transform of $x_{2,k,l}(t)$.
Since $x_{k,l}(t)$ is an infinite pulse train it has infinite bandwidth. The bandwidth $B'$ of each Zak-OTFS carrier signal (and that of the Zak-OTFS subframe) is simply the bandwidth of the delay domain factor $w_1(\tau)$ and hence the delay axis spread of $w_1(\tau)$ is approximately $1/B'$.

Therefore, the delay and Doppler spread of the pulse shaping filter $w_{tx}(\tau, \nu)$ are inversely related to the bandwidth and time of the Zak-OTFS subframe. Hence a larger bandwidth and time implies smaller DD domain spread of $w_{tx}(\tau, \nu)$ which in turn implies a smaller spread of the effective channel filter $h_{\mbox{\scriptsize{eff}}}(\tau, \nu)$ (see (\ref{eqnhefftwisted})). A smaller spread implies that the crystallization condition in (\ref{eqncryscnd}) can be satisfied for a physical channel with higher delay and Doppler shifts.


The sinc pulse shaping filter is given by
\begin{eqnarray}
w_{tx}(\tau, \nu) & = & \sqrt{BT} \, sinc(B \tau) \, sinc(T \nu).
\end{eqnarray}
For the sinc filter, the Zak-OTFS subframe duration $T' = T$ and the subframe bandwidth $B' = B$, so the spectral efficiency is $BT/(B' T') = 1$ symbol/dimension. The root raised cosine (RRC) filter is given by
\begin{eqnarray}
\label{eqnp4rrc1}
    w_{tx}(\tau, \nu) & = & \sqrt{BT} \, rrc_{\beta_{\tau}}(B \tau) \, rrc_{\beta_{\nu}}(\nu T),
\end{eqnarray}where $0 \leq \beta_{\tau}, \beta_{\nu} \leq 1$ and
\begin{eqnarray}
\label{eqnp4rrc2}
rrc_{_{\beta}}(x) & \hspace{-3mm} = & \hspace{-3mm} \frac{\sin(\pi x (1 - \beta)) + 4 \beta x \cos(\pi x (1 + \beta))}{\pi x \left( 1 - (4 \beta x)^2 \right)}.
\end{eqnarray}
The Zak-OTFS subframe duration $T' = T(1 + \beta_{\nu})$
and the subframe bandwidth $B' = B(1 + \beta_{\tau})$, so the
spectral efficiency is $BT/(B' T') = \frac{1}{(1 + \beta_{\tau})(1 + \beta_{\nu})} < 1$ symbol/dimension.

The RRC function decays faster than the sinc function with the rate of decay controlled by the roll-off parameter $\beta$. Note that the sinc and RRC filters coincide when $\beta = 0$. Faster decay leads to less DD domain aliasing and greater predictability (for more details see Figs. $4$ and $5$ from \cite{zakotfs2}). The cost is higher subframe duration and subframe bandwidth,
and reduced spectral efficiency.

This motivates the next Section where we introduce Gaussian filters with better localization than sinc and RRC filters.

    \vspace{-3mm}
    \section{Zak-OTFS with Gaussian pulse}
    \label{sec3}
    Given $B = M \nu_p$ and $T = N \tau_p$, the unit energy Gaussian pulse shaping filter is given by
    \begin{eqnarray}\label{eqn_w1gp}
        w_{tx}(\tau,\nu) & =  & w_1(\tau) \, w_2(\nu),
        \end{eqnarray}where the factors  $w_1(\tau)$ and $w_2(\nu)$ are given by
\begin{eqnarray}
\label{eqn_w1gp1}
        w_1(\tau) & \Define & \left( \frac{2 \alpha_{\tau} B^2}{\pi} \right)^{\frac{1}{4}} \, e^{- \alpha_{\tau} B^2 \tau^2} \nonumber \\
         w_2(\nu) & \Define & \left( \frac{2 \alpha_{\nu} T^2}{\pi} \right)^{\frac{1}{4}} \, e^{- \alpha_{\nu} T^2 \nu^2}. 
    \end{eqnarray}Since Gaussian pulses have infinite support, we work with the time interval $T'$ where $99 \%$ of the Zak-OTFS subframe energy is localized in the TD, and with the bandwidth $B'$ where $99 \%$ of the Zak-OTFS subframe energy is localized in the FD. We configure the Gaussian pulse by adjusting the parameters $\alpha_{\tau}$ and $\alpha_{\nu}$. For example, the case of no time/bandwidth expansion ($T' = T$ and $B' = B$) corresponds to setting $\alpha_{\tau} = \alpha_{\nu} = 1.584$.

\begin{figure}
\hspace{-2mm}
\includegraphics[width=9.5cm, height=6.4cm]{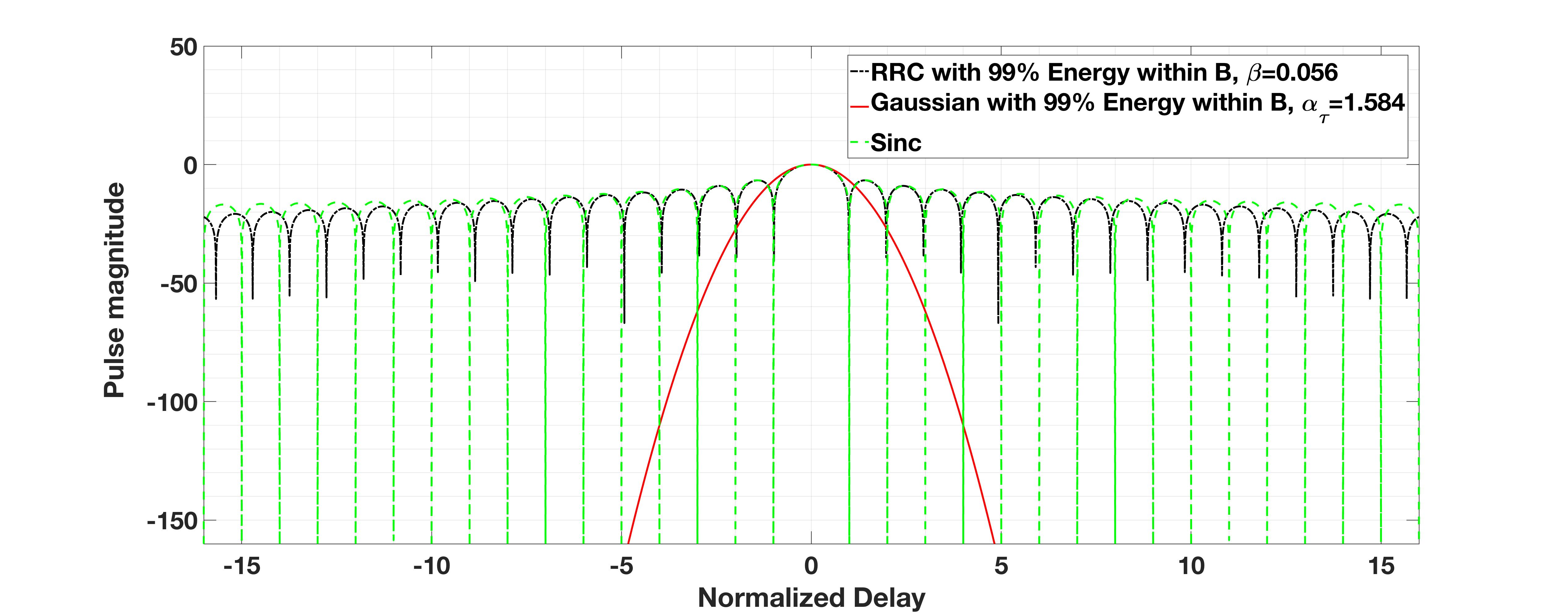}
\caption{Pulse magnitude $10 \log_{10}\vert w(\tau) \vert$
in dB as a function of the normalized delay $B \tau$.
Gaussian filters (red) exhibit better localization than sinc filters (green) and RRC filters (black) in the case of no time/bandwidth expansion.}
\label{figgaussianp}
\end{figure}

   We compare the Gaussian pulse shaping filter with the RRC filter given by (\ref{eqnp4rrc1}), (\ref{eqnp4rrc2}). Here the case of no time/bandwidth expansion corresponds to the roll-off parameter $\beta = 0.056$. The RRC function decays faster than the sinc function leading to less DD domain aliasing and greater predictability (for more details see Fig.~$4$ and $5$ from \cite{zakotfs2}).

   Fig.~\ref{figgaussianp} illustrates the advantage of Gaussian filters over sinc and RRC filters in the case of no time/bandwidth expansion. Outside the main lobe, the magnitude of the Gaussian pulse is significantly lower, leading to less DD domain aliasing and greater predictability. 


At the receiver we apply a pulse shaping filter $w_{rx}(\tau, \nu)$ that is matched to the transmit filter
    \begin{eqnarray}
\label{eqn_wrx}
    w_{rx}(\tau, \nu) & = & e^{j 2 \pi \nu \tau} \, w_{tx}^*(-\tau, -\nu).
\end{eqnarray}
It follows from (\ref{eqn_w1gp}), (\ref{eqn_w1gp1}) that the matched Gaussian pulse shaping filter is given by

{\vspace{-4mm}
\small
    \begin{eqnarray}\label{eqn_w2gp}
        w_{rx}(\tau,\nu) & = &  \left( \frac{2 \alpha_{\tau} B^2}{\pi} \right)^{\frac{1}{4}} \, e^{- \alpha_{\tau} B^2 \tau^2} \, \, e^{j 2 \pi \nu \tau } \nonumber \\
        & & \left( \frac{2 \alpha_{\nu} T^2}{\pi} \right)^{\frac{1}{4}} \, e^{- \alpha_{\nu} T^2 \nu^2}.
    \end{eqnarray}\normalsize}
It is remarkable that by working with Gaussian filters in the DD domain, we are able to derive closed form expressions for the effective channel filter and the covariance of the noise. 
    \begin{theorem}
    \label{paper3thm1}
      For the Gaussian transmit pulse shaping filter in (\ref{eqn_w1gp}) and the corresponding receive match filter in
      (\ref{eqn_w2gp}), the effective discrete DD domain channel filter is given by
    \vspace{-3mm}
    \begin{align}
    \label{eqn925330}
        h_{\mbox{\scriptsize{eff}}}[k,l]                  =&
                                    \sum\limits_{i=1}^{P}h_i
                                    e^{-\frac{1}{2}\big(\alpha_{\tau}B^2(\tau_i - \frac{k\tau_p}{M})^2+\alpha_{\nu}T^2(\nu_i - \frac{l\nu_p}{N})^2\big)} \nonumber \\
                                    &e^{-\frac{\pi^2}{2}\big(\frac{\nu_i^2}{\alpha_{\tau}B^2}+\frac{k^2\tau_p^2}{M^2\alpha_{\nu}T^2}\big)}
                                    e^{-j\pi(\tau_i\nu_i-\frac{kl}{NM})}.
    \end{align}
    \end{theorem}
\begin{IEEEproof}
    See Appendix \ref{apx_heff}.
\end{IEEEproof}

    \begin{theorem}
    \label{p3thm2}
        For all $k_1, k_2=0,1,\cdots, M-1$, $l_1, l_2 = 0,1 , \cdots, N-1 $, the $(k_1N+l_1+1,k_2N+l_2+1)$-th element of the covariance matrix of the DD domain noise vector ${\bf n}_{\mbox{\scriptsize{dd}}}$ in (\ref{eqn87624}) is
        $\mathbb{E}[n_{\mbox{\scriptsize{dd}}}[k_1,l_1]n_{\mbox{\scriptsize{dd}}}^*[k_2,l_2]]$, whose expression is given by (\ref{eqn8366401}) (see top of next page). 
    \end{theorem}
    \begin{IEEEproof}
See Appendix \ref{apx_cov}.
    \end{IEEEproof}
           \begin{figure}
     \hspace{-1mm} 
        \includegraphics[width=9cm,height=6.5cm]{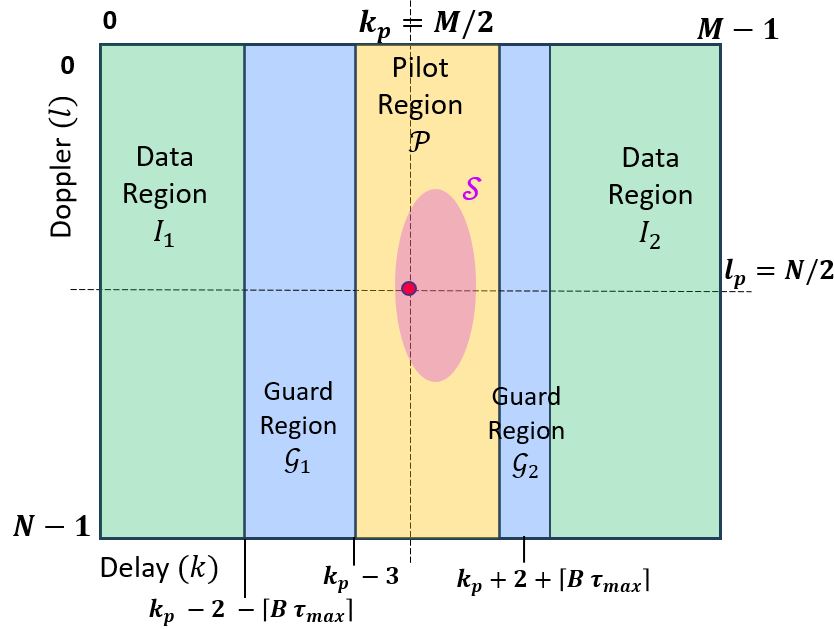}
        \vspace{-4mm}
        \caption{The fundamental region ${\mathcal D}_0$ of the Zak-OTFS modulation tiled with pilot, guard and data regions. The pilot is located at $(k_p, l_p) = (M/2, N/2)$ and the information symbols modulate Zak-OTFS carriers located in the data regions. The pilot region contains carriers located in the region ${\mathcal S} + (k_p, l_p)$ where ${\mathcal S}$ is the support of the effective channel. It extends to the left of the pilot location illustrating a physical channel with a zero delay path and a shaping filter that leaks energy outside the main lobe.} 
        \label{fig_subframe}
    \end{figure}  
    \begin{figure*}[!t]
        \vspace{-8mm}
        {\small
        \begin{eqnarray}
        \label{eqn8366401}
                 \mathbb{E}[n_{\mbox{\scriptsize{dd}}}[k_1,l_1]n_{\mbox{\scriptsize{dd}}}^*[k_2,l_2]] & \hspace{-2.5mm} = &  \hspace{-3mm} N_0 \frac{\tau_p}{T} \sqrt{\frac{2 \pi }{\alpha_{\nu}}} \sum\limits_{q_1 = - \infty}^{\infty} \sum\limits_{q_2 = - \infty}^{\infty}  e^{-j 2 \pi \frac{q_1 l_1 - q_2 l_2}{N}} \, e^{-\frac{\pi^2 \tau_p^2}{\alpha_{\nu} T^2} \left( \left( q_1 + \frac{k_1}{M} \right)^2 +  \left( q_2 + \frac{k_2}{M} \right)^2  \right)} \, e^{- \frac{\alpha_{\tau} B^2}{2} \left( \frac{(k_2 - k_1) \tau_p}{M} + (q_2 - q_1) \tau_p \right)^2}.
        \end{eqnarray}\normalsize}   
        \vspace{-2mm}
        \hrulefill        
        \vspace{-4mm}
     \end{figure*}
Since the Gaussian transmit pulse shaping filter $w_{tx}(\tau, \nu)$ has unit energy, the average energy of the transmitted Zak-OTFS modulated signal $s_{\mbox{\scriptsize{td}}}(t)$ is equal to the average energy of the corresponding
discrete DD domain signal $x_{\mbox{\scriptsize{dd}}}[k,l]$, i.e.

{\vspace{-4mm}
\small
\begin{eqnarray}
    {\mathbb E}\left[ \int \left\vert s_{\mbox{\scriptsize{td}}}(t) \right\vert^2 \, dt \right] & = & \sum\limits_{k=0}^{M-1} \sum\limits_{l=0}^{N-1} {\mathbb E}\left[ \left\vert x_{\mbox{\scriptsize{dd}}}[k,l] \right\vert^2 \right].
\end{eqnarray}\normalsize}
    \subsection{Integration of pilot and data}
     We transmit a pilot at location $(k_p, l_p)$ and we surround the pilot with a region (Pilot Region ${\mathcal P}$ + Guard Region ${\mathcal G} = {\mathcal G}_1 \, \cup \, {\mathcal G}_2$ in Fig.~\ref{fig_subframe}) where no data is transmitted. The number of information symbols transmitted is $\vert {\mathcal I} \vert$ where $\vert {\mathcal I} \vert$ denotes the number of carriers
     contained in ${\mathcal I} = {\mathcal I}_1 \, \cup \, {\mathcal I}_2$. For $0 \leq k < M$ and $0 \leq l < N$,
     the symbol $x[k,l]$ (see Section \ref{sec2apaper4}) is given by
    \begin{eqnarray}
      x[k,l] & = \begin{cases}
       \sqrt{\frac{E_d}{\vert {\mathcal I} \vert}} \, x_I[k,l] &, (k,l) \in  {\mathcal I} \\
          \sqrt{E_p}  &, (k,l) = (k_p, l_p) \\
          0  &, \mbox{\scriptsize{otherwise}} \\
      \end{cases}
    \end{eqnarray}where $x_I[k,l]$ is the information symbol
    transmitted on the $(k,l)$-th carrier. Given that ${\mathbb E}\left[ \left\vert x_I[k,l] \right\vert^2 \right]= 1$, the average energy transmitted in a Zak-OTFS subframe is $E_p + E_d$ and the average transmit power is $(E_p + E_d)/T'$.

    We normalize the path gains by taking ${\mathbb E}\left[ \vert h_i \vert^2 \right] = 1$ so that the received power
    is equal to the transmit power. At the receiver, the noise power is $N_0 B'$, hence the ratio of data power to noise power is
    \begin{eqnarray}
    \gamma_d & \Define & \frac{E_d}{N_0 B' T'},
    \end{eqnarray} and the ratio of pilot power to noise power is
    \begin{eqnarray}
    \gamma_p & \Define & \frac{E_p}{N_0 B' T'}.
    \end{eqnarray}
   We refer to $E_p/E_d = \gamma_p/\gamma_d$ as the ratio of pilot power to data power (PDR).

    In each Zak-OTFS subframe, the effective discrete DD domain channel filter $h_{\mbox{\scriptsize{eff}}}[k,l]$ is estimated based
        on the received pilot symbols at DD locations within the pilot region ${\mathcal P}$ (this estimation is same as the model-free estimation in $(30)$ of \cite{zakotfs2}). This estimate is then used to construct an estimate of the effective channel matrix ${\bf H}_{\mbox{\scriptsize{dd}}}$ (see (\ref{eqn87624})). The received DD symbols $y_{\mbox{\scriptsize{dd}}}[k,l]$, $(k,l) \in \left( {\mathcal I} \, \cup {\mathcal G} \right)$ are used to detect the transmitted information symbols. These received $(MN - \vert {\mathcal P} \vert)$ symbols are arranged into a vector which is simply the vector of $\vert {\mathcal I} \vert $ transmitted information symbols times the effective channel matrix plus the noise vector.\footnote{\footnotesize{Since information symbols are not transmitted in the pilot region, the effective channel matrix is a sub-matrix of ${\bf H}_{\mbox{\scriptsize{dd}}}$ in (\ref{eqn87624}) where we retain only those columns which correspond to the elements of the transmit vector ${\bf x}_{\mbox{\scriptsize{dd}}}$ where information symbols are embedded. Similarly, we retain only those rows of ${\bf H}_{\mbox{\scriptsize{dd}}}$ which correspond to elements of the received vector ${\bf y}_{\mbox{\scriptsize{dd}}}$ which do not lie in the pilot region.}} An MMSE equalizer is used to detect the information symbols from the vector of received DD domain symbols in ${\mathcal I} \cup {\mathcal G}$.
    
                \begin{figure}
        \includegraphics[width=9.0cm,height=6.5cm]{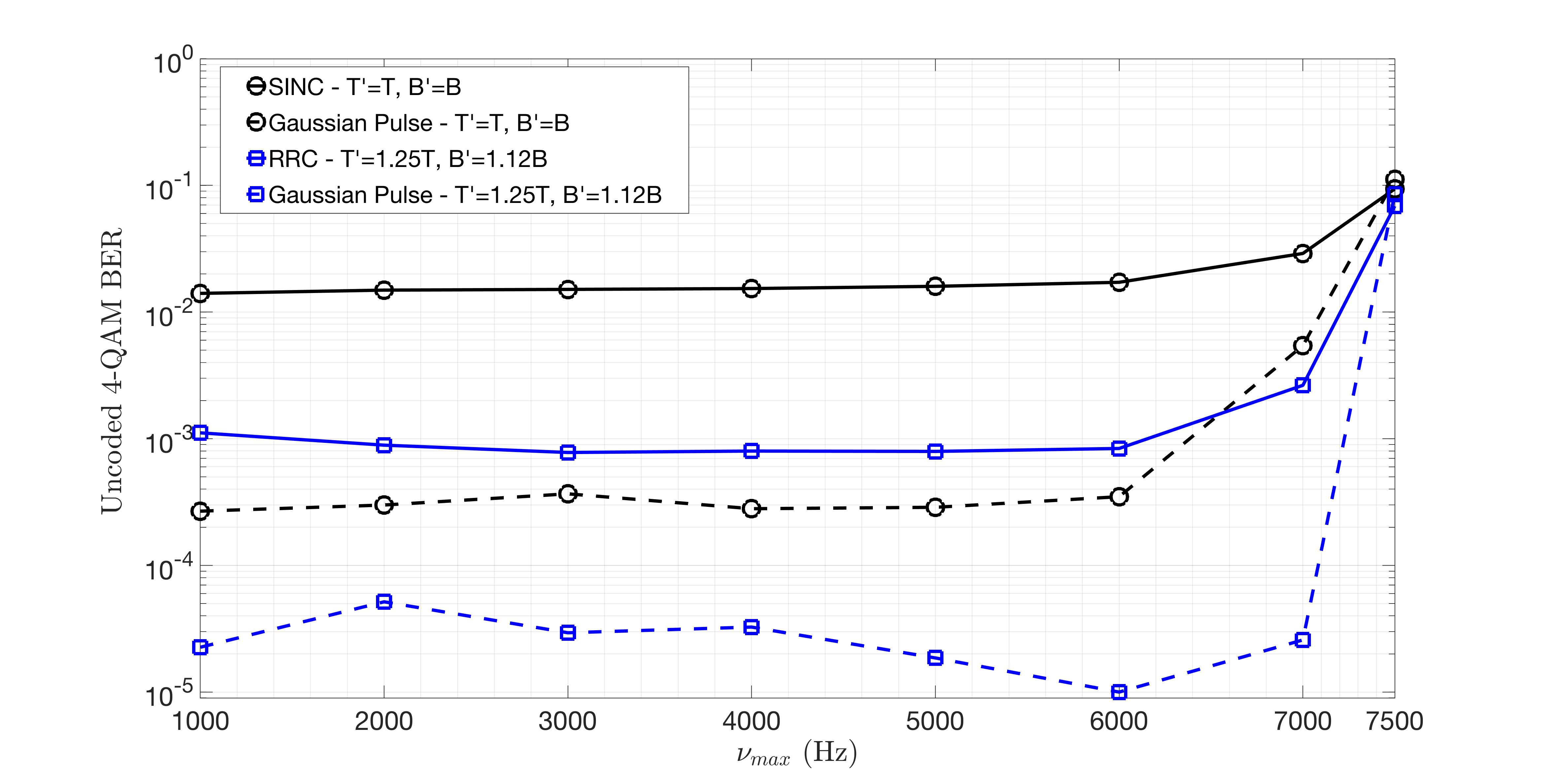}
        \caption{Uncoded $4$-QAM BER vs. $\nu_{max}$. Data SNR $\gamma_d = 25$ dB and PDR $\gamma_p/\gamma_d = 5$ dB. Black curves illustrate BER performance when there is no time/bandwidth expansion. Blue curves illustrate BER performance when $B' = 1.12B$ and $T' = 1.25T$.}
        \label{fig_numax_ber}
        \vspace{-4mm}
    \end{figure}

    \section{Numerical Results}    
   \label{simsecpaper4}
   In this Section we measure uncoded $4$-QAM BER performance of Zak-OTFS for different pulse shaping waveforms (sinc, RRC, and Gaussian). Throughout this Section we fix $M=32$, $N=48$, $\nu_p = 15$ KHz (unless stated otherwise),
   and we fix the pilot location $(k_p, l_p)=(M/2, N/2 )$ as shown in Fig.~\ref{fig_subframe}. The bandwidth $B = M \nu_p = 480$ KHz
   and the duration $T = N \tau_p = 3.2$ ms. Also throughout this Section, we consider the six-path Veh-A channel model \cite{EVAITU} with power-delay profile listed in Table \ref{tab_veha}. The relative power $p_i$ of the $i$-th path is the ratio (in dB) of the mean squared value of the $i$-th tap to that of the first tap ($p_i = {\mathbb E}\left[ \vert h_i \vert^2 \right]/{\mathbb E}\left[ \vert h_1 \vert^2 \right]$, $i=1,2,\cdots, 6$). We then normalize these mean squared values so that $\sum\limits_{i=1}^6 {\mathbb E}\left[ \left\vert h_i \right\vert^2 \right] = 1$. The Doppler shift $\nu_i$ induced by the $i$-th channel path is modeled as $\nu_{max} \cos(\theta_i)$ where $\nu_{max}$ is the maximum Doppler shift and $\theta_i, i=1,2,\cdots, 6$ are modeled as i.i.d. random variables distributed uniformly in $[0 \,,\, 2 \pi)$.


       \begin{figure}
        \includegraphics[width=9cm,height=6.5cm]{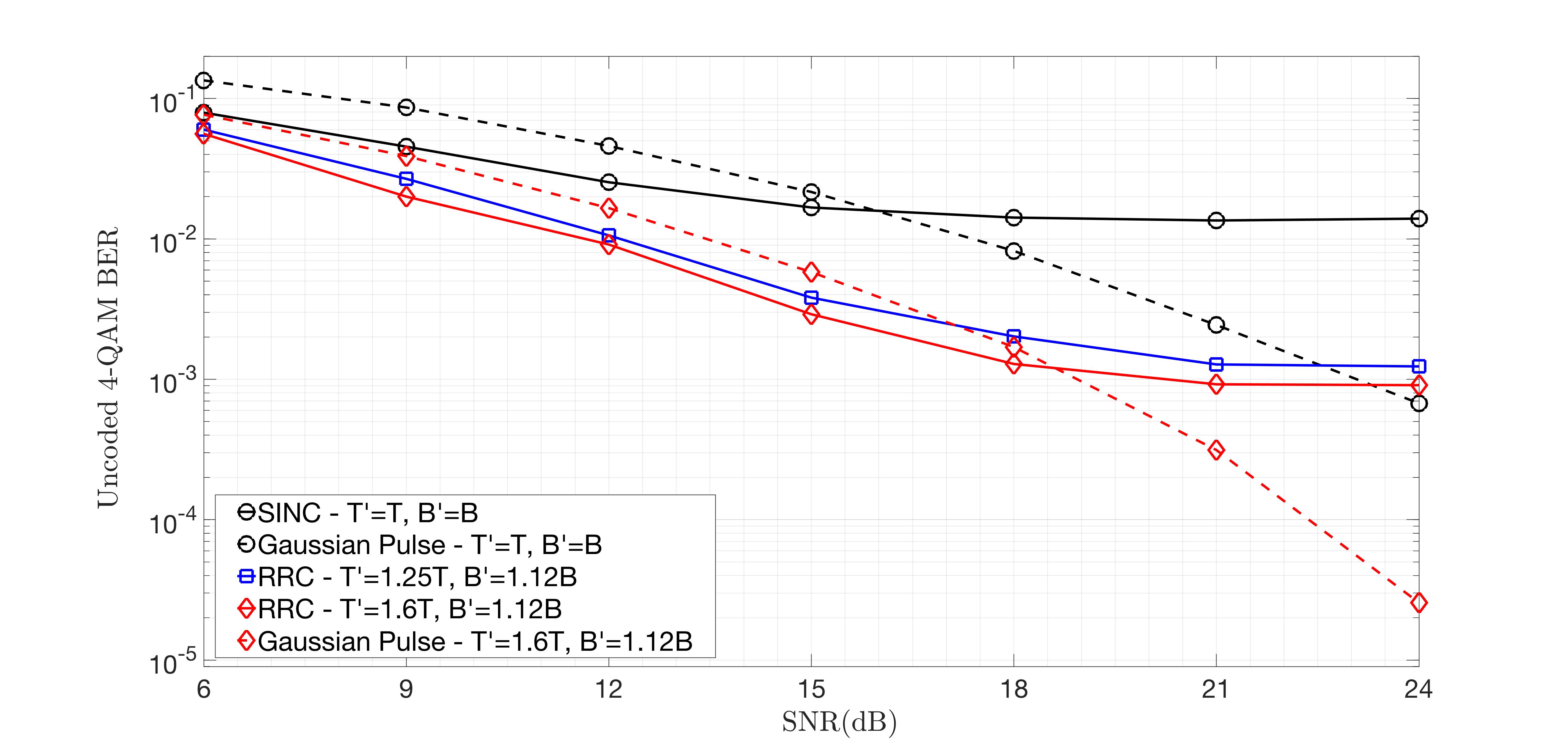}
        \vspace{-2mm}
        \caption{Uncoded $4$-QAM BER as a function of increasing SNR $\gamma_d$. PDR $\gamma_p/\gamma_d = 5$ dB and $\nu_{max} = 815$ Hz. Black curves illustrate BER performance when there is no time/bandwidth expansion. Blue curves illustrate BER performance
        when $B' = 1.12B$ and $T' = 1.25T$. Red curves illustrate BER performance when $B' = 1.12B$ and $T' = 1.6T$. Solid curves correspond to RRC/sinc filters and dashed curves correspond to Gaussian filters. BER performance with Gaussian filters does not saturate at high SNR.}
        \label{fig_snr_ber}
        \vspace{-2mm}
    \end{figure}       
    
Fig.~\ref{fig_numax_ber} plots BER of uncoded $4$-QAM as a function of increasing Doppler spread ($2 \nu_{max}$). When there is no time/bandwidth expansion ($B' = B, T' = T$) the RRC filter coincides with the sinc filter. When $\nu_{max} < 6$ KHz, the crystallization conditions in (\ref{eqncryscnd}) hold, and BER performance of the Gaussian filter (black dashed curve) is significantly better than that of the sinc filter (black solid curve). Selecting the roll-off factor $\beta_{\tau} = 0.12$ for the RRC filter reduces delay domain aliasing at the cost of bandwidth expansion ($B' = 1.12B$). Choosing a roll-off factor $\beta_{\nu} = 0.25$ reduces Doppler domain aliasing at the cost of expanding time duration ($T' = 1.25T$).
Simultaneous time and bandwidth expansion ($T' = 1.25T$ and $B' = 1.12B$)
improves BER performance for the RRC filter (blue solid curve). We remark that when we increased $\beta_{\nu}$ to expand the time duration to $T' = 1.9T$ we observed negligible improvement in BER performance, and we have not shown this curve in Fig.~\ref{fig_numax_ber}. However, a Gaussian pulse designed for $B' = 1.12B$ and $T' = 1.25T$ (blue dashed curve) improves BER performance significantly.

Fig.~\ref{fig_snr_ber} plots BER of uncoded $4$-QAM as a function of increasing SNR $\gamma_d$, for a fixed $\nu_{max} = 815$ Hz, and for a fixed PDR $\gamma_p/\gamma_d = 5$ dB. Regardless of time/bandwidth expansion, BER performance of RRC filters (solid curves) saturates at high SNR, whereas BER performance of Gaussian filters (dashed curves) does not. This is because the RRC filter leaks more energy outside the main lobe than the Gaussian filter. Since PDR $\gamma_p/\gamma_d$ is fixed, the pilot power $\gamma_p$ increases with the SNR $\gamma_d$ and BER performance with RRC/sinc filters is limited at high SNR by interference from the pilot to the data symbols.  
  
       \begin{figure}
      
        \includegraphics[width=9cm,height=6.5cm]{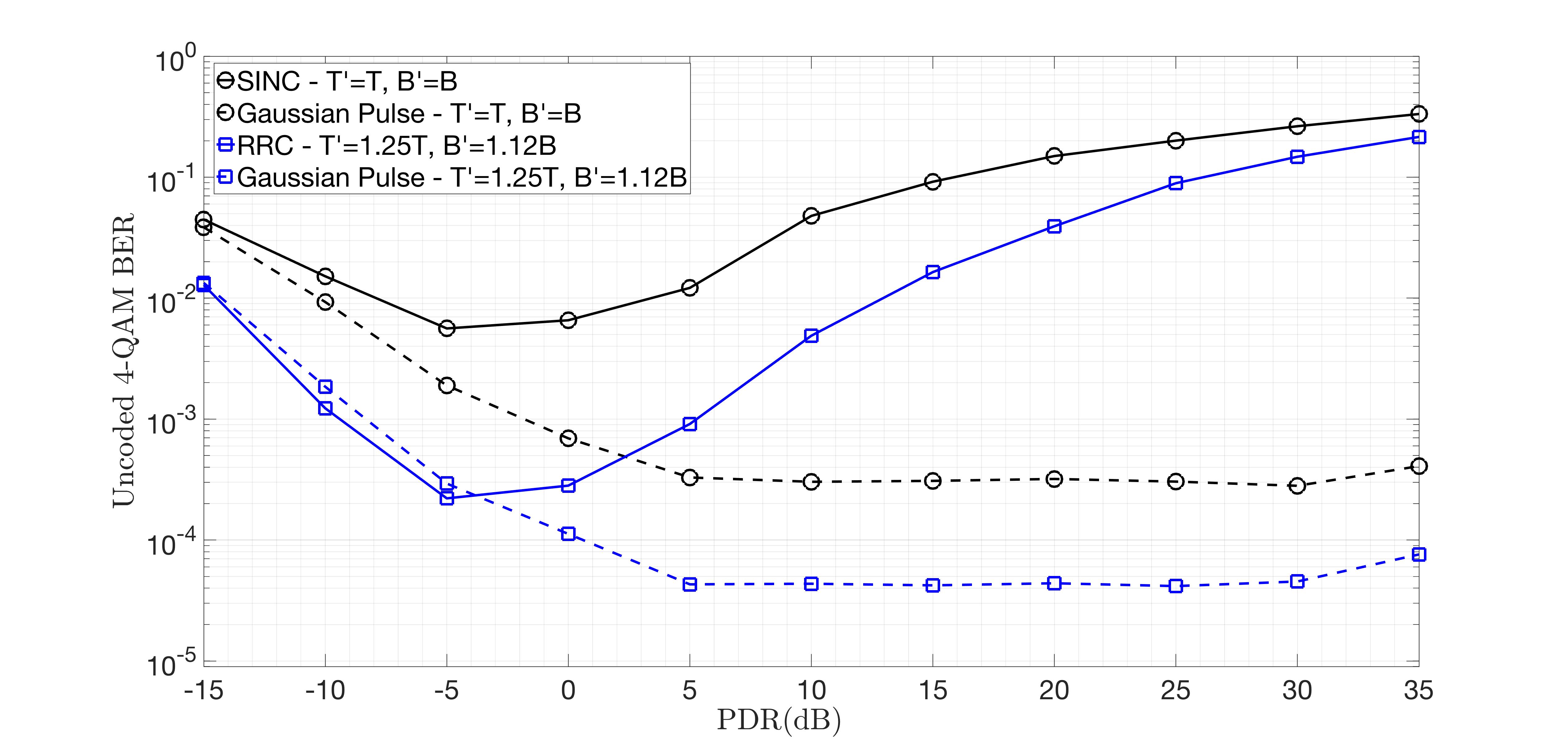}
        \caption{Uncoded $4$-QAM BER as a function of increasing PDR. SNR $\gamma_d = 25$ dB and $\nu_{max} = 815$ Hz. Solid curves indicate BER performance for RRC filters and dashed curves indicate BER performance for Gaussian filters. Black curves illustrate BER performance when there is no time/bandwidth expansion, blue curves illustrate BER performance when $B'= 1.12B$ and $T' = 1.25T$. The characteristic ``U" shape of the solid curves results from RRC filters leaking significant energy into the data region.} 
        \label{fig_pdr_ber}
        \vspace{-5mm}
    \end{figure}  

Fig.~\ref{fig_pdr_ber} plots BER of uncoded $4$-QAM as a function of increasing PDR for a fixed $\nu_{max} = 815$ Hz and for a fixed SNR $\gamma_d = 25$ dB. The pilot used for channel
sensing leaks energy into the data region and the leakage
interferes with the carriers used for data transmission. Energy leakage with Gaussian filters (dashed curves) is small, whereas energy leakage with RRC filters (solid curves) is more significant. When interference is small (i.e., low PDR), BER performance improves with increasing PDR as the channel estimate becomes more accurate. With RRC filters, when the pilot to noise ratio $\gamma_p$ exceeds the data SNR $\gamma_d$ (PDR $> 0$ dB) interference becomes more significant than noise and BER degrades with increasing PDR. This explains the characteristic ``U" shape of the solid curves. 
    
            \begin{figure}
     \hspace{-2mm} 
        \includegraphics[width=9.5cm,height=6.0cm]{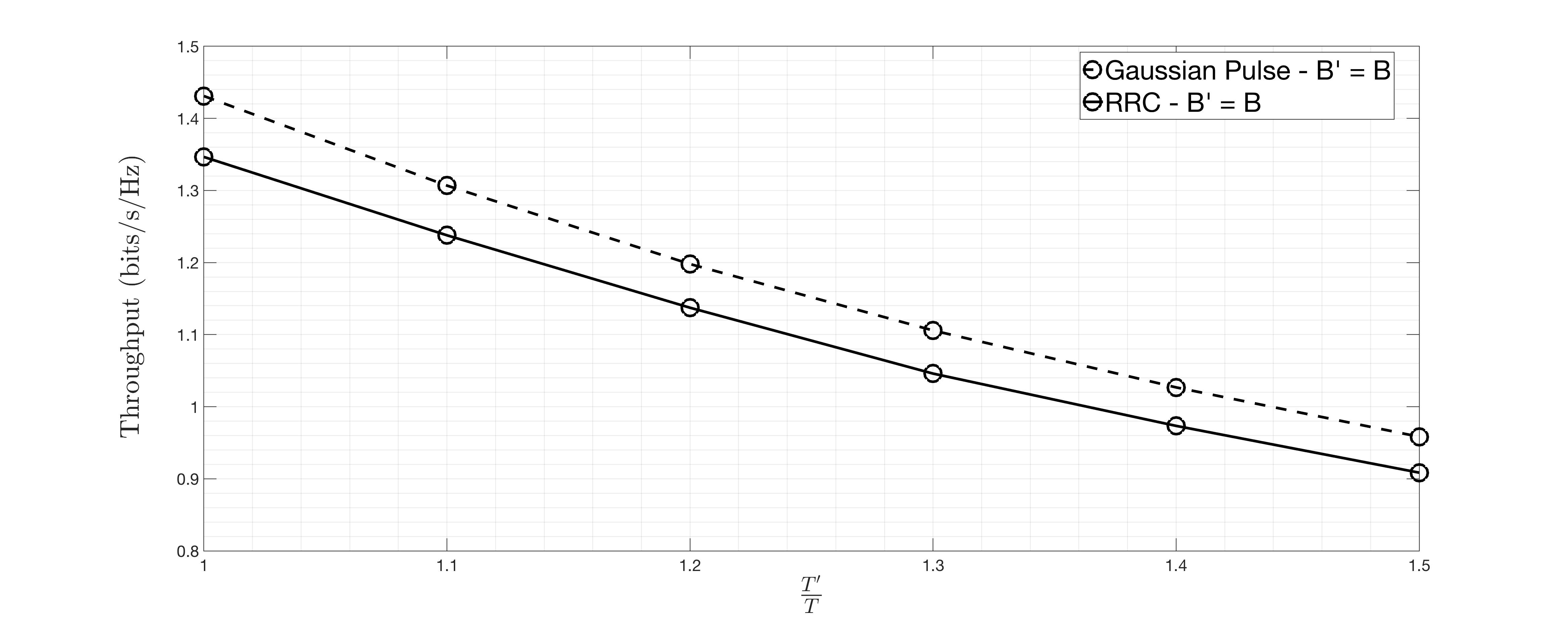}
        \vspace{-2mm}
        \caption{Effective throughput (bits/sec/Hz) as a function of time expansion $T'/T$. No bandwidth expansion ($B' = B$), PDR $\gamma_p/\gamma_d = 5$ dB, SNR $\gamma_d = 25$ dB, and $\nu_{max} = 6$ KHz. Gaussian filters (dashed curve) deliver higher effective throughput than RRC filters (solid curve).}
        \label{fig_thput}
        \vspace{-5mm}
    \end{figure} 

  Fig.~\ref{fig_thput} plots effective throughput as a function of time expansion $T'/T$. There is no bandwidth expansion ($B' = B$), PDR $\gamma_p/\gamma_d = 5$ dB, SNR $\gamma_d = 25$ dB and $\nu_{max} = 6$ KHz. Effective throughput is the ratio of the average number of bits reliably communicated in each subframe to the available
  degrees of freedom $(B' T')$. The average number of information bits reliably communicated is $2 \vert {\mathcal I} \vert (1 - H(BER))$ where $\vert {\mathcal I} \vert $ is the number of Zak-OTFS carriers in the data region ${\mathcal I}$ and $H$ denotes the binary entropy function. We observe that Gaussian filters deliver higher
  effective throughput than RRC filters.

  When the crystallization conditions are satisfied, the Zak-OTFS I/O relation is predictable and non-fading. However, the crystallization conditions apply to the effective channel and the pulse shaping filter contributes to the effective channel. When the Doppler spread $2 \nu_{max}$ is high and the crystallization condition along the Doppler domain is satisfied ($\nu_p > \nu_{ds} > 2 \nu_{max}$, see(\ref{eqncryscnd})), the delay period $\tau_p = 1/\nu_p < 1/(2 \nu_{max})$ must be small. When the pulse shaping filter leaks energy outside the main lobe, delay-domain aliasing leads to frequency selectivity as shown in Fig.~\ref{fig_aliasing}. By limiting leakage we limit the support of the effective channel and extend that part of the hyperbola $\tau_p \, \nu_p = 1$ for which I/O relation is predictable and non-fading.

  We now quantify the impact of pulse shaping on I/O predictability by modifying the Veh-A channel model considered hitherto. We suppose that the product $\tau_{max} \, \nu_{max} = 0.1$ so that the points $(\tau_{max} , \nu_{max})$ lie on a hyperbola below the period curve $\tau_p \, \nu_p = 1$.
 We can interpret filter leakage as shifting the lower hyperbola towards the upper hyperbola. We scale the delays of the six channel paths listed in Table \ref{tab_veha} by
 $\tau_{max}/2.51$ and we model the path Doppler shifts in the same way as before.

\begin{figure}
     \hspace{-2mm} 
        \includegraphics[width=10cm,height=6.5cm]{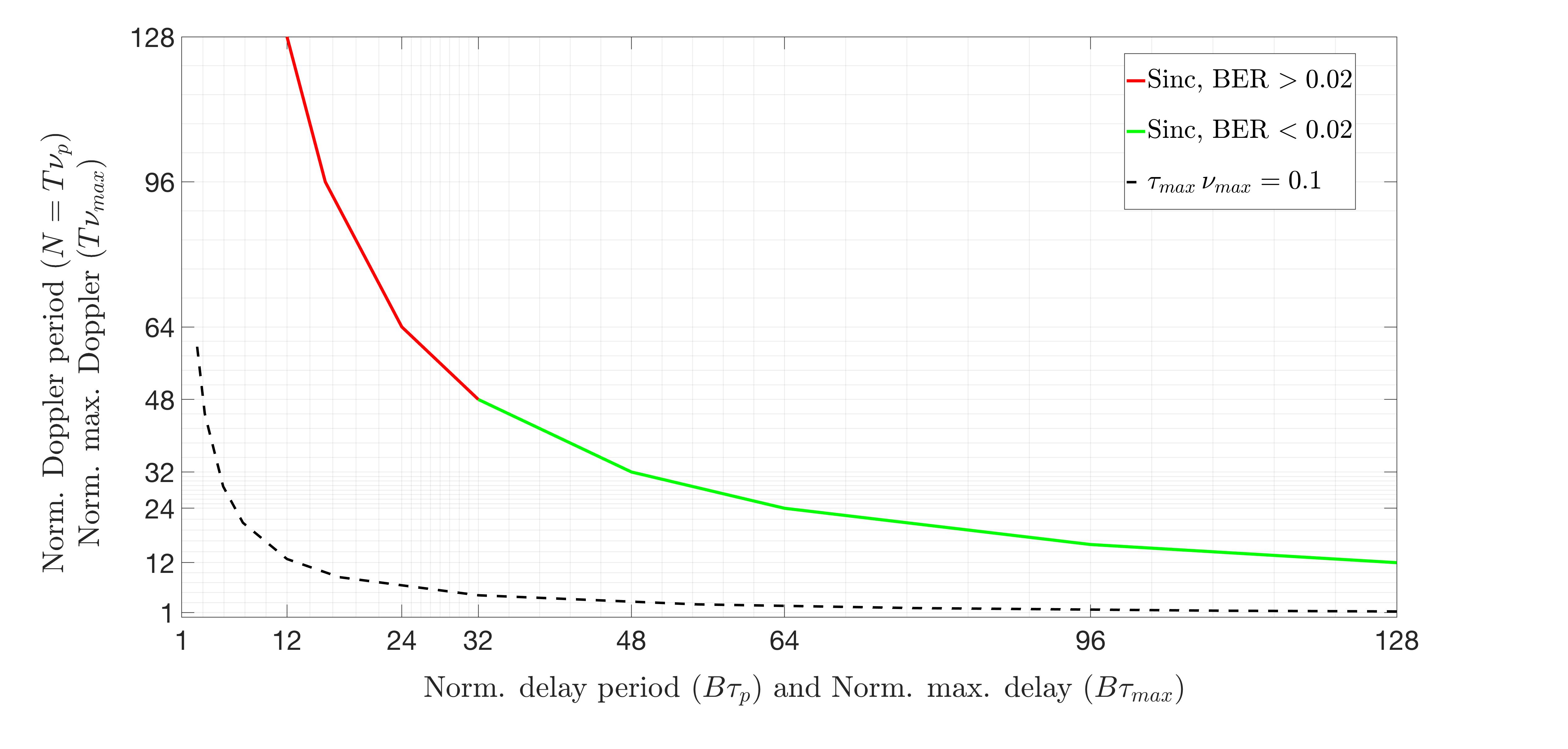}
        \vspace{-2mm}
        \caption{Reliable communication (green curve) with the sinc pulse. The crystallization conditions hold and reliable communication is defined to be BER $< 0.02$. The points $(\tau_{max}, \nu_{max})$ lie on the hyperbola $\tau_{max} \, \nu_{max} = 0.1$. }
        \label{fig_hyperbola_sinc}
        \vspace{-5mm}
    \end{figure}

\begin{figure}
     \hspace{-2mm} 
        \includegraphics[width=10cm,height=6.3cm]{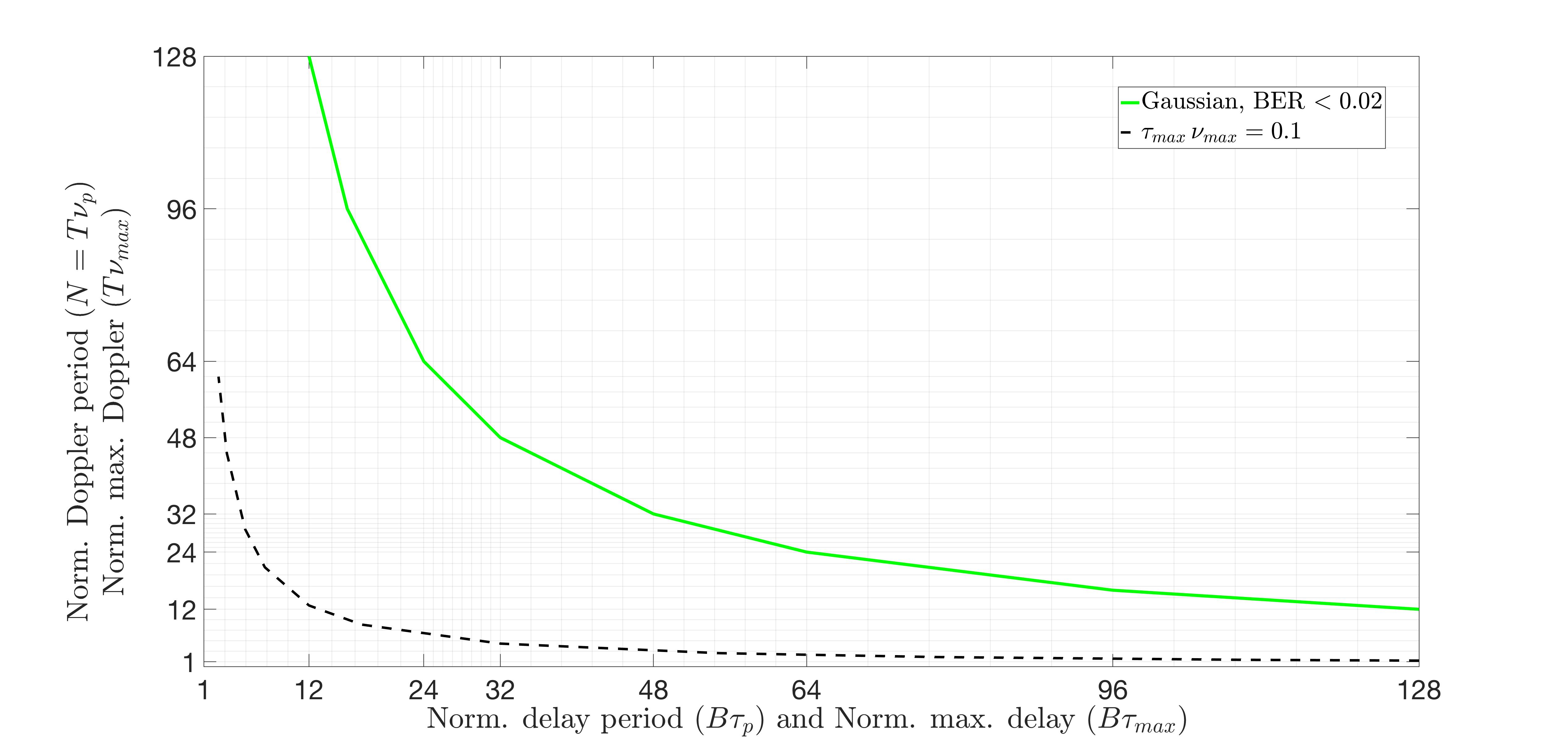}
        \vspace{-2mm}
        \caption{Reliable communication (green curve) with the Gaussian pulse. The crystallization conditions hold and reliable communication is defined to be BER $< 0.02$. The points $(\tau_{max}, \nu_{max})$ lie on the hyperbola $\tau_{max} \, \nu_{max} = 0.1$.}
        \label{fig_hyperbola_Gaussian}
        \vspace{-5mm}
    \end{figure}

 We suppose there is no time/bandwidth expansion and we set $B' = B = 0.48$ MHz and $T' = T = 3.2$ ms. We restrict our attention to points $(\tau_p, \nu_p)$ on the hyperbola $\tau_p \, \nu_p = 1$ for which $M = B \tau_p$ and $N = T \nu_p$ are both integers having product $MN = BT = 1536$, and for which it is possible to transmit at least one information bit within the subframe structure described in Fig.~\ref{fig_subframe}. Specifically, we consider $(M,N) = (128,12), (96, 16), (64, 24)$, $(48,32), (32,48), (24, 64)$, $(16, 96)$ and $(12, 128)$. For each of these pairs $(M,N)$ we
 simulate the uncoded $4$-QAM BER for $\nu_p = B/M$, $\tau_p = 1/\nu_p = T/N$, $\nu_{max} = \frac{\nu_p}{2} - 1000$ Hz, and $\tau_{max} = 0.1/\nu_{max}$. Note that the Doppler domain crystallization condition $\nu_p > 2 \nu_{max}$ is satisfied for infinite $(B,T)$. When BER $< 0.02$ 
 we consider communication to be reliable, and we color the hyperbola $\tau_p \, \nu_p = 1$ green, otherwise we color it red (see Fig.~\ref{fig_hyperbola_sinc} and Fig.~\ref{fig_hyperbola_Gaussian} for the sinc and Gaussian filter respectively). From these figures it is clear that the region of predictable operation (green segment of the hyperbola $\tau_p \, \nu_p = 1$) is larger for the Gaussian filter than for the sinc filter. Table \ref{tab_predictop} lists all parameters for the eight simulation points $(M,N)$ and indicates for each point whether or not the sinc and Gaussian filters support reliable communication. In the Table, while all choices of $(M,N)$ satisfy the conditions $\tau_p > \tau_{max}$ and $\nu_p > 2 \nu_{max}$ (crystallization conditions with infinite $(B,T)$, i.e., no pulse shaping), with finite $(B,T)$ the crystallization condition in (\ref{eqncryscnd}) is not satisfied for the sinc filter for three out of the eight cases whereas it is satisfied in all cases for the Gaussian filter. 
 

        \begin{table*}[]
        \vspace{-6mm}
            \centering
            \caption{Predictability at operating points on the hyperbola $(B \tau_p) (T \nu_p) = 1536$.}
            \begin{tabular}{|c|c|c|c|c|c|c|c|c|}
                \hline
                S.No.       &  $M$ &  $N$  & $\nu_p$ &  $\tau_p$      &  $\nu_{max}$  &  $\tau_{max}$  & Sinc pulse & Gaussian pulse   \\
                &  & & (in KHz) & (in $\mu s$) &  (in KHz) & (in $\mu s$)  & Predictable & Predictable \\
                &  &  &  &  &  & & Operation  & Operation \\
                \hline
                $1$. & $128$ &  $12$ &  $3.75$ & $266.6$  &  $0.875$ & $114.3$ & \textcolor{green}{\checkmark}  &   \textcolor{green}{\checkmark} \\
                \hline
                 $2$. & $96$ &  $16$ &  $5$ & $200$  &  $1.5$ & $66.7$ & \textcolor{green}{\checkmark}  &   \textcolor{green}{\checkmark} \\
                \hline
                                 $3$. & $64$ &  $24$ &  $7.5$ & $133.3$  &  $2.75$ & $36.4$ & \textcolor{green}{\checkmark}  &   \textcolor{green}{\checkmark} \\
                \hline
         $4$. & $48$ &  $32$ &  $10$ & $100$  &  $4$ & $25$ & \textcolor{green}{\checkmark}  &   \textcolor{green}{\checkmark} \\
                \hline
         $5$. & $32$ &  $48$ &  $15$ & $66.66$  &  $6.5$ & $15.4$ & \textcolor{green}{\checkmark}  &   \textcolor{green}{\checkmark} \\
                \hline
     $6$. & $24$ &  $64$ &  $20$ & $50$  &  $9$ & $11.1$ & \textcolor{red}{$\times$}  &   \textcolor{green}{\checkmark} \\
                \hline
     $7$. & $16$ &  $96$ &  $30$ & $33.33$  &  $14$ & $7.1$ & \textcolor{red}{$\times$}  &   \textcolor{green}{\checkmark} \\
                \hline
   $8$. & $12$ &  $128$ &  $40$ & $25$  &  $19$ & $5.3$ & \textcolor{red}{$\times$}  &   \textcolor{green}{\checkmark} \\
                \hline
            \end{tabular}
            \label{tab_predictop}
        \end{table*}
        

    \section{Conclusions}
We have introduced a framework for filter design where the symplectic Fourier transform connects aliasing in the DD domain with time/bandwidth expansion. Predictability of the Zak-OTFS I/O relation results from minimizing aliasing in the DD domain, so our framework provides insight into the time-bandwidth expansion required to introduce AI/ML methods in 6G wireless systems. We have introduced Gaussian filters in the DD domain and have demonstrated that they support integrated sensing and communication within a single Zak-OTFS subframe. Our method is to transmit a pilot in the center of a subframe, and to surround the pilot region with a guard region to mitigate interference between pilot and data. The choice of transmit filter determines the fraction of pilot energy that lies outside the pilot region, and the degradation in BER that results from the pilot interference to data carriers. We have demonstrated that Gaussian filters improve BER performance compared to standard choices such as sinc and RRC filters.

    \appendices

    \begin{figure*}[!t]
        \vspace{-5mm}
        \begin{eqnarray}
        \label{eqn936416}
                 \mathbb{E}[n_{\mbox{\scriptsize{dd}}}[k_1,l_1]n_{\mbox{\scriptsize{dd}}}^*[k_2,l_2]] &= & \frac{2 B \tau_p}{T} \sqrt{\frac{\alpha_{\tau}}{\alpha_{\nu}}} \sum\limits_{q1 = -\infty}^{\infty} \sum\limits_{q2 = -\infty}^{\infty} {\Bigg [} e^{j 2 \pi \frac{q_2 l_2 - q_1 l_1}{N}} \, e^{-\frac{\pi^2 \tau_p^2}{\alpha_{\nu} T^2} \left( \left( q_1 + \frac{k_1}{M} \right)^2 +  \left( q_2 + \frac{k_2}{M} \right)^2  \right)} \, \nonumber \\
                 & & \hspace{43mm} {\mathbb E}\left[  v\left( q_1 \tau_p + \frac{k_1 \tau_p}{M} \right) \, v^*\left( q_2 \tau_p + \frac{k_2 \tau_p}{M} \right) \right] \, {\Bigg ]}
        \end{eqnarray}        
        \vspace{-2mm}
        \hrulefill        
        \vspace{-1mm}
     \end{figure*}
    \begin{figure*}[!t]
        \vspace{-1mm}
        {\small
        \begin{eqnarray}
        \label{eqn026559}
                 {\mathbb E}\left[  v\left( q_1 \tau_p + \frac{k_1 \tau_p}{M} \right) \, v^*\left( q_2 \tau_p + \frac{k_2 \tau_p}{M} \right) \right] & \hspace{-2.5mm} = & \hspace{-3mm} \iint e^{- \alpha_{\tau} B^2 (\tau_1^2 + \tau_2^2)} \, \underbrace{{\mathbb E}\left[ n\left( \frac{k_1 \tau_p}{M} + q_1 \tau_p - \tau_1 \right) \, n^*\left( \frac{k_2 \tau_p}{M} + q_2 \tau_p - \tau_2 \right)\right]}_{= N_0 \delta\left( \tau_2 - \tau_1 - \frac{(k_2 - k_1)\tau_p}{M} - (q_2 - q_1)\tau_p \right), \,\, \mbox{\scriptsize{since $n(t)$ is AWGN}}} \, d\tau_1 \, d\tau_2 \nonumber \\
                 & \hspace{-2.5mm} = & \hspace{-3mm} N_0 \int e^{- \alpha_{\tau} B^2 \left( \tau_1^2 + \left( \tau_1 + \frac{(k_2 - k_1) \tau_p}{M} + (q_2 - q_1) \tau_p\right)^2 \right)} \, d\tau_1 \nonumber \\
                 & \hspace{-2.5mm} = & \hspace{-3mm} N_0 \sqrt{\frac{\pi }{2 \alpha_{\tau} B^2}} \, e^{- \frac{\alpha_{\tau} B^2}{2} \left( \frac{(k_2 - k_1) \tau_p}{M} + (q_2 - q_1) \tau_p \right)^2}
        \end{eqnarray}\normalsize}   
        \vspace{-2mm}
        \hrulefill        
        \vspace{-4mm}
     \end{figure*}

        \section{Proof of Theorem \ref{paper3thm1}}
         \label{apx_heff}
        Since twisted convolution is associative, it follows from (\ref{eqn13p5}) that
        \begin{eqnarray}
        \label{eqn97535}
            h_{\mbox{\scriptsize{eff}}}(\tau, \nu) & = & w_{rx}(\tau, \nu) *_{\sigma} {\Bigg (} h_{\mbox{\scriptsize{phy}}}(\tau, \nu) *_{\sigma} w_{tx}(\tau, \nu) {\Bigg )}. \nonumber \\
        \end{eqnarray}The expression for $h_{\mbox{\scriptsize{phy}}}(\tau, \nu)$ given by (\ref{eqnp2}) yields
        \begin{eqnarray}
        \label{eqn42p4556}
            h_{\mbox{\scriptsize{phy}}}(\tau, \nu) *_{\sigma} w_{tx}(\tau, \nu) & & \nonumber \\
            & & \hspace{-30mm} = \sum\limits_{i=1}^P h_i w_1(\tau - \tau_i) \, w_2(\nu - \nu_i) \, e^{j 2 \pi \nu_i (\tau - \tau_i)}.
        \end{eqnarray}Substituting (\ref{eqn42p4556}) in (\ref{eqn97535}) gives
        \begin{eqnarray}
        \label{eqn65356}
            h_{\mbox{\scriptsize{eff}}}(\tau, \nu) & & \nonumber \\
            & & \hspace{-20mm} =  \sum\limits_{i=1}^P h_i \, w_{rx}(\tau, \nu) *_{\sigma} {\Bigg (} w_1(\tau - \tau_i) \, w_2(\nu - \nu_i) \, e^{j 2 \pi \nu_i (\tau - \tau_i)} {\Bigg )}  \nonumber \\
            & & \hspace{-20mm} \mya \sum\limits_{i=1}^P h_i \, e^{j 2 \pi \nu_i (\tau - \tau_i)} \iint {\Bigg [} w_1(\tau') w_2(\nu') w_1(\tau - \tau' - \tau_i) \,  \nonumber \\
            & & \hspace{1mm} e^{j 2 \pi \nu' \tau} \, w_2(\nu - \nu' - \nu_i) \, e^{-j 2 \pi \nu_i \tau'}   {\Bigg ]} \, d\tau' \, d\nu'
        \end{eqnarray}where step (a) follows from substituting for $w_{rx}(\tau, \nu)$ using (\ref{eqn_wrx}). We substitute for $w_1(\cdot)$ and $w_2(\cdot)$ using (\ref{eqn_w1gp}) and solve the integrals to obtain
        \begin{eqnarray}
            h_{\mbox{\scriptsize{eff}}}(\tau, \nu) & \hspace{-3mm} = & 
            \hspace{-3mm} \sum\limits_{i=1}^P h_i \, e^{j \pi (\nu \tau - \nu_i \tau_i)} \, e^{- \frac{\alpha_{\tau} B^2 (\tau - \tau_i)^2}{2}} \, e^{- \frac{\alpha_{\nu} T^2 (\nu - \nu_i)^2}{2}} \nonumber \\
            & & \hspace{5mm} e^{- \frac{\pi^2 \nu_i^2}{2 \alpha_{\tau} B^2}} \,  e^{- \frac{\pi^2 \tau^2}{2 \alpha_{\nu} T^2}}.
        \end{eqnarray}Since $h_{\mbox{\scriptsize{eff}}}[k,l]$ is given by sampling $h_{\mbox{\scriptsize{eff}}}(\tau, \nu)$ on the information grid/lattice $\Lambda_{\mbox{\scriptsize{dd}}}$ (see \ref{eqn13p5}), we obtain (\ref{eqn925330}).

\section{Proof of Theorem \ref{p3thm2}}\label{apx_cov}
            The Zak transform of AWGN $n(t)$ is given by
            \begin{align}
            n_{\mbox{\scriptsize{dd}}}(\tau,\nu) &= \sqrt{\tau_p}\sum\limits_{q=-\infty}^{\infty} n(\tau +q\tau_p)e^{-j2\pi \nu q\tau_p}.
            \end{align} 
            The noise at the output of the receive pulse shaping filter is given by
            \begin{eqnarray}
            \label{eqn_nzak}
                n_{\mbox{\scriptsize{dd}}}^{w_{rx}}(\tau,\nu) & = & w_{rx}(\tau,\nu) *_{\sigma} n_{\mbox{\scriptsize{dd}}}(\tau,\nu) \nonumber \\
                & & \hspace{-26mm} = \sqrt{\frac{2 B \tau_p}{T}} \, \left( \frac{\alpha_{\tau}}{\alpha_{\nu}} \right)^{\frac{1}{4}}
                                                        \hspace{-2mm} \sum\limits_{q=-\infty}^{\infty} 
                                                        \hspace{-2mm}                
                                                        e^{-j2\pi\nu q\tau_p}
                                                        v(\tau+q\tau_p)  \, e^{- \frac{\pi^2 \left( q \tau_p + \tau \right)^2}{\alpha_{\nu} T^2}}  \nonumber \\
                v(\tau) & \Define & \int e^{- \alpha_{\tau} B^2 \tau'^2 } \, n(\tau - \tau') \, d\tau'. 
            \end{eqnarray}Sampling $n_{\mbox{\scriptsize{dd}}}^{w_{rx}}(\tau,\nu)$ at $(\tau=\frac{k\tau_p}{M},\nu=\frac{l\nu_p}{N})$ gives
            
             {\vspace{-4mm}
             \small
             \begin{align}\label{eqn_nsampled}
              \hspace{-3mm}n_{\mbox{\scriptsize{dd}}}[k,l]\hspace{-1mm}=& \hspace{-0.5mm} \sqrt{\frac{2 B \tau_p}{T}} \, \left( \frac{\alpha_{\tau}}{\alpha_{\nu}} \right)^{\frac{1}{4}}
                                                        \hspace{-2mm}
                                                \sum\limits_{q=-\infty}^{\infty} \hspace{-1.5mm}
                                                e^{-j2\pi \tfrac{l q}{N}}                        
                                                e^{\tfrac{-\pi^2\tau_p^2}{\alpha_{\nu}T^2}\left(\tfrac{k}{M}+q\right)^2} \hspace{-2mm}
                                               v(\tfrac{k\tau_p}{M}\hspace{-1mm}+\hspace{-1mm}q\tau_p)\hspace{-4mm}
             \end{align}\normalsize}For any $k_1, k_2 = 0,1,\cdots, M-1$, $l_1, l_2 = 0,1 \cdots, N-1$, we calculate the covariance ${\mathbb E}\left[ n_{\mbox{\scriptsize{dd}}}[k_1,l_1] \,  n_{\mbox{\scriptsize{dd}}}^*[k_2,l_2] \right]$ using (\ref{eqn_nsampled}) to obtain (\ref{eqn936416}) (see top of this page).
             Using the expression for $v(\tau)$ given by (\ref{eqn_nzak}), the expectation in the RHS of (\ref{eqn936416}) is given by (\ref{eqn026559}).
             Substituting (\ref{eqn026559}) in (\ref{eqn936416})
             yields the expression for the covariance given by (\ref{eqn8366401}). 

\section{Derivation of (\ref{paper4carrierwaveforms})}
\label{appCp4}
From (\ref{eqn3p4}), (\ref{qpeqn}) and (\ref{eqn2}) it follows that
\begin{eqnarray}
    x_{\mbox{\scriptsize{dd}}}(\tau, \nu) & \Define & \sum\limits_{k=0}^{M-1} \sum\limits_{l=0}^{N-1} x[k,l] \, x_{\mbox{\scriptsize{dd}},k,l}(\tau, \nu)
\end{eqnarray}where

{\vspace{-4mm}
\small
\begin{eqnarray}
     x_{\mbox{\scriptsize{dd}},k,l}(\tau, \nu ) & \hspace{-3.5mm} \Define & \hspace{-6mm} \sum\limits_{n,m \in {\mathbb Z}}  \hspace{-1.5mm}  e^{j 2 \pi \frac{n l}{N}} \, \delta\left(\tau - n\tau_p - \frac{k \tau_p}{M}\right) \delta\left(\nu - m \nu_p - \frac{l \nu_p}{N}\right), \nonumber \\
\end{eqnarray}\normalsize}$k=0,1,\cdots, M-1$, $l=0,1,\cdots,N-1$, is the quasi-periodic DD pulse localized at $\left(\frac{k \tau_p}{M}, \frac{l \nu_p}{N}\right)$ in the continuous DD domain.
Since twisted convolution is a linear operation
\begin{eqnarray}
\label{eqn_xddwtxkl}
    x_{\mbox{\scriptsize{dd}}}^{w_{tx}}(\tau, \nu) & = & w_{tx}(\tau, \nu) \, *_{\sigma}  \,  x_{\mbox{\scriptsize{dd}}}(\tau, \nu) \nonumber \\
    & = & \sum\limits_{k=0}^{M-1} \sum\limits_{l=0}^{N-1} x[k,l] \, x_{\mbox{\scriptsize{dd}},k,l}^{w_{tx}}(\tau, \nu ) ,\nonumber \\
    x_{\mbox{\scriptsize{dd}},k,l}^{w_{tx}}(\tau, \nu ) & \Define & w_{tx}(\tau, \nu) \, *_{\sigma} \, x_{\mbox{\scriptsize{dd}},k,l}(\tau, \nu ).
\end{eqnarray}Therefore, the transmit TD signal in (\ref{eqn8p4}) is given by
\begin{eqnarray}
\label{eqn53p4}
    s_{\mbox{\scriptsize{td}}}(t) & = & {\mathcal Z_t}^{-1}{\Big (}  x_{\mbox{\scriptsize{dd}}}^{w_{tx}}(\tau , \nu ) {\Big )} \nonumber \\
    & = & \sum\limits_{k=0}^{M-1} \sum\limits_{l=0}^{N-1} x[k,l] \, {\mathcal Z}_t^{-1}\left(x_{\mbox{\scriptsize{dd}},k,l}^{w_{tx}}(\tau, \nu ) \, \right) \nonumber \\
    & = & \sum\limits_{k=0}^{M-1} \sum\limits_{l=0}^{N-1} x[k,l] \,  s_{\mbox{\scriptsize{td}},k,l}(t), \nonumber \\
    s_{\mbox{\scriptsize{td}},k,l}(t) & \Define & {\mathcal Z_t}^{-1}{\Big (}  x_{\mbox{\scriptsize{dd}},k,l}^{w_{tx}}(\tau , \nu ) {\Big )},
\end{eqnarray}where the second step follows from (\ref{eqn_xddwtxkl}) and the linearity of the Inverse Zak transform.
Note that, $s_{\mbox{\scriptsize{td}},k,l}(t)$ is the carrier waveform for the $(k,l)$-th information symbol $x[k,l]$.
Let
\begin{eqnarray}
\label{eqn_22_1}
x_{k,l}(t) & \hspace{-3mm} \Define &  \hspace{-3mm} {\mathcal Z}_t^{-1} {\Big (} x_{\mbox{\scriptsize{dd}},k,l}(\tau, \nu ) {\Big )} \nonumber \\
& &  \hspace{-6mm} = {\mathcal Z}_t^{-1} {\Bigg (} \sum\limits_{n,m \in {\mathbb Z}}  \hspace{-1.5mm}  e^{j 2 \pi \frac{n l}{N}} \, \delta\left(\tau - n\tau_p - \frac{k \tau_p}{M}\right) \nonumber \\
& &  \hspace{19mm} \delta\left(\nu - m \nu_p - \frac{l \nu_p}{N}\right) {\Bigg )}.
\end{eqnarray}
Next, we show that the Zak transform of the TD signal $\iint w_{tx}(\tau', \nu') \, x_{k,l}(t - \tau' )  \, e^{j 2 \pi \nu' (t - \tau')} \, d\tau \, d\nu$ is $x_{\mbox{\scriptsize{dd}},k,l}^{w_{tx}}(\tau , \nu )$ (see (\ref{eqnprf2}) at the top of this page).
\begin{figure*}
\vspace{-5mm}
\begin{eqnarray}
\label{eqnprf2}
    {\mathcal Z}_t \left( \iint w_{tx}(\tau', \nu') \, x_{k,l}(t - \tau' )  \, e^{j 2 \pi \nu' (t - \tau')} \, d\tau' \, d\nu' \right) & = & \nonumber \\
    & & \hspace{-60mm} = \sqrt{\tau_p} \iint \sum\limits_{n \in {\mathbb Z}} w_{tx}(\tau', \nu') \, x_{k,l}(\tau + n \tau_p - \tau' )  \, e^{j 2 \pi \nu' (\tau + n \tau_p - \tau')} \, e^{-j 2 \pi n \nu \tau_p} \, d\tau' \, d\nu' \nonumber \\
    & & \hspace{-60mm} \mya  \iint w_{tx}(\tau', \nu') \, \underbrace{\left[ \sqrt{\tau_p} \sum\limits_{n \in {\mathbb Z}}  x_{k,l}(\tau + n \tau_p - \tau' )  \, e^{-j 2 \pi (\nu - \nu')  n \tau_p}  \right]}_{= {\mathcal Z}_t\left( x_{k,l}(t) \right) \, \mbox{\small{at}} \, (\tau - \tau', \nu - \nu')} \, e^{j 2 \pi \nu' (\tau - \tau')} \, d\tau' \, d\nu' \nonumber \\
    & & \hspace{-60mm} \myb \iint w_{tx}(\tau', \nu') \, x_{\mbox{\scriptsize{dd}},k,l}(\tau - \tau', \nu - \nu' ) \, e^{j 2 \pi \nu' (\tau - \tau')} \, d\tau' \, d\nu' \,\, = \,\, w_{tx}(\tau, \nu) \, *_{\sigma} \, x_{\mbox{\scriptsize{dd}},k,l}(\tau, \nu) \nonumber \\
    & & \hspace{-60mm} \myc x_{\mbox{\scriptsize{dd}},k,l}^{w_{tx}}(\tau, \nu).
\end{eqnarray}
\begin{eqnarray*}
\hline 
\end{eqnarray*}
\end{figure*}
In (\ref{eqnprf2}), steps (a) and (b) follow from the definition of Zak transform of a TD signal and the fact that $x_{\mbox{\scriptsize{dd}},k,l}(\tau, \nu)$ is the Zak transform of $x_{k,l}(t)$ (see (\ref{eqn_22_1}), also see Eqn.~$(4)$ in \cite{zakotfs1} for details on Zak transform).
Step (c) follows from (\ref{eqn_xddwtxkl}).
Comparing (\ref{eqn53p4}) with (\ref{eqnprf2}) it then follows that
\begin{eqnarray}
s_{\mbox{\scriptsize{td}}, k,l}(t) & \hspace{-3mm} = &  \hspace{-3mm}  {\mathcal Z_t}^{-1}{\Big (}  x_{\mbox{\scriptsize{dd}},k,l}^{w_{tx}}(\tau , \nu ) {\Big )} \nonumber \\
&  & \hspace{-15mm}  = \iint w_{tx}(\tau, \nu) \, x_{k,l}(t - \tau )  \, e^{j 2 \pi \nu (t - \tau)} \, d\tau \, d\nu.
\end{eqnarray}
We apply the integral expression for the inverse Zak transform to the DD domain signal in (\ref{eqn_22_1}) to obtain (see Eqn.~$(7)$ in \cite{zakotfs1} for details on inverse Zak transform)
\begin{eqnarray}
x_{k,l}(t) & \hspace{-3mm} = &  \hspace{-3mm} {\mathcal Z}_t^{-1} {\Big (} x_{\mbox{\scriptsize{dd}},k,l}(\tau, \nu ) {\Big )} \nonumber \\ 
&  &  \hspace{-20mm} = \sqrt{\tau_p} \, \int\limits_{0}^{\nu_p} x_{\mbox{\scriptsize{dd}},k,l}(t, \nu ) \, d\nu  \nonumber \\
& & \hspace{-20mm} =  \sqrt{\tau_p} \sum\limits_{n,m \in {\mathbb Z}} {\Bigg [} e^{j 2 \pi \frac{n l}{N}} \delta\left(t - n\tau_p - \frac{k \tau_p}{M}\right) \, \nonumber \\
& & \int\limits_{0}^{\nu_p} \delta\left(\nu - m \nu_p - \frac{l \nu_p}{N}\right) \, d\nu {\Bigg ]} \nonumber \\
&  &  \hspace{-20mm} = \sqrt{\tau_p} \, \sum\limits_{n \in {\mathbb Z} } e^{j 2 \pi \frac{n l }{N} } \, \delta\left(t - n\tau_p - \frac{k \tau_p}{M}\right),
\end{eqnarray}since
\begin{eqnarray}
    \int\limits_{0}^{\nu_p} \delta\left(\nu - m \nu_p - \frac{l \nu_p}{N}\right) \, d\nu  & = \begin{cases}
        0 &, m \ne 0 \\
        1 &, m = 0
    \end{cases}.
\end{eqnarray}


\end{document}